\documentclass[twocolumn,aps,prd,reprint,showpacs,amsmath,amssymb,nofootinbib]{revtex4-1}

\usepackage{graphicx} 
\usepackage{enumerate}
\usepackage{subfigure}
\usepackage{dcolumn} 
\usepackage{bm} 
\usepackage{hyperref} 
\usepackage{epsfig}
\usepackage{color}
\usepackage{soul}

\def\ba{\begin{eqnarray}}
\def\ea{\end{eqnarray}}
\def\bea{\begin{eqnarray}}
\def\eea{\end{eqnarray}}
\def\be{\begin{equation}}
\def\ee{\end{equation}}
\def\d{\mathrm{d}}
\def\({\left(}
\def\){\right)}
\def\[{\left[}
\def\]{\right]}

\begin{document}

\preprint{IGC-17/8-1}

\title{A cosmological open quantum system}

\author{Sarah Shandera$^{1}$}
\email{shandera@gravity.psu.edu}
\author{Nishant Agarwal$^{\P, 2, 3}$} 
\email{nishant\_agarwal@uml.edu}
\author{Archana Kamal$^{\P, 2, 4}$}
\email{archana\_kamal@uml.edu}
\thanks{\\ $^{\P}$ These authors contributed equally to this work.}
\affiliation{
$^{1}$Institute for Gravitation and the Cosmos, The Pennsylvania State University, University Park, PA 16802, USA \\
$^{2}$Department of Physics and Applied Physics, University of Massachusetts, Lowell, MA 01854, USA \\
$^{3}$\mbox{Lowell Center for Space Science and Technology, University of Massachusetts, Lowell, MA 01854, USA} \\
$^{4}$Research Laboratory of Electronics, Massachusetts Institute of Technology, Cambridge, MA 02139, USA}

\date{\today}

\begin{abstract}
We derive the evolution equation for the density matrix of a UV- and IR- limited band of comoving momentum modes of the canonically normalized scalar degree of freedom in two examples of nearly de Sitter universes. Including the effects of a cubic interaction term from the gravitational action and tracing out a set of longer wavelength modes, we find that the evolution of the system is non-Hamiltonian and non-Markovian. We find linear dissipation terms for a few modes with wavelength near the boundary between system and bath, and nonlinear dissipation terms for all modes. The non-Hamiltonian terms in the evolution equation persist to late times when the scalar field dynamics is such that the curvature perturbation continues to evolve on super-Hubble scales. 
\end{abstract}

\pacs{Valid PACS appear here}

\maketitle


\section{Introduction}
\label{sec:intro}

In cosmological models for the primordial universe, the unavoidable quantum fluctuations of matter and of the linearized gravitational field are the original source of the rich structure of late-time inhomogeneities observed today as variations in the temperature of the cosmic microwave background (CMB) \cite{Ade:2015xua} and the distribution of galaxies \cite{Alam:2016hwk}. Two goals in cosmology are to use the classical data collected from the CMB and large-scale structure to pinpoint the particle physics of the primordial era, and to understand whether signatures of their quantum origin may remain observable today.

A major focus of inflation model building in the last decade or so has been the study of how particle interactions during or just after inflation may generate non-Gaussianity in the correlation functions of the inhomogeneities. Optimistically, this interest was fueled by the notion that information from statistics beyond the power spectrum could eventually distinguish among the zoo of particle physics mechanisms for inflation. However, the fact that our cosmological observations today are limited to a finite volume of space, leaving sufficiently long wavelength modes fundamentally unobservable, leads to an interesting conundrum for studies of inflationary particle physics via post-inflation statistics: if fluctuations with wavelengths observable to us can be coupled to fluctuations on unobservable scales, there is additional non-Gaussian sample variance \cite{Bartolo:2012sd,Nelson:2012sb,LoVerde:2013xka,Nurmi:2013xv,Thorsrud:2013kya,Thorsrud:2013mma,Byrnes:2013ysj,Loverde2014,Bonga:2015urq,Deutsch:2017rsn} that affects the precision with which inferences can be made from the data. In non-single clock inflation scenarios (roughly, models where more than one light degree of freedom contributes to the fluctuations) this cosmic variance uncertainty can be equal to or larger than current observational uncertainty.

Cosmic variance from mode-coupling is a statistical phenomena at the level of classical correlators, and can be calculated on any constant time slice after inflation. However, within the inflationary paradigm, those statistics are generated in the course of a dynamical, quantum mechanical process. Considering the full quantum story of inflation models that couple modes of different wavelengths may lead to additional insight into the nature of information contained in the inflationary fluctuations. Here, we use the fact of IR-limited observational cosmology and the associated issue of classical non-Gaussian cosmic variance, as motivation to investigate the quantum evolution equations of a system of cosmological modes coupled to a bath of these long wavelength modes, during an inflationary era. For simplicity, we focus our attention on a single cubic interaction term from the Einstein-Hilbert action that, depending on a choice for the scalar field dynamics, can support long-short mode-coupling. 

The role of gravity in this scenario is three-fold: (1) the homogeneous, isotropic, time-dependent gravitational background serves as a zero-momentum pump sourcing pairs of quanta in two-mode squeezed states; the zero-momentum nature of the pump ensures a homogeneous and isotropic amplification of all the momentum modes $k \leq aH$, (2) the inherently nonlinear gravitational action itself provides the coupling term between system and bath, and (3) the cosmological horizon of an observer after the inflationary, quasi-de Sitter era puts a long wavelength limit to the observable modes and so forces the longer wavelength modes into the (unobservable) bath.

Our approach is complementary to previous works on open systems in inflation \cite{Burgess:2014eoa, Burgess:2015ajz, Boyanovsky:2015tba, Boyanovsky:2015jen,Nelson:2016kjm,Hollowood:2017bil,Martin:2018zbe,Martin:2018lin} which have so far considered the opposite case of computing the evolution for super-Hubble, long wavelength modes in a bath of sub-Hubble, short wavelength modes. By tracing out the long-wavelength modes instead, we give a fully quantum treatment of observables which remains valid even in the presence of a strong coupling between long- and short-wavelength modes, i.e. when long wavelength modes cannot be absorbed by a renormalization of the background clock. In the semi-classical limit our results should recover not only the mean late-time curvature correlators that are usually calculated, but also the full super-cosmic variance probability distributions for how classical statistics observed in a single Hubble volume may differ from the mean statistics of the model \cite{Nelson:2012sb,LoVerde:2013xka}.

The paper is organized as follows. We briefly review the Hamiltonian for fluctuations in quasi de Sitter space in section\ \ref{sec:model} and describe the evolution equations in the two example scenarios we consider (slow-roll and non-attractor) in section\ \ref{sec:srna}. These two background evolutions, and the choice of a particular interaction term, allow us the simplest possible calculation to examine the difference between models with and without coupling between modes of very different wavelengths (system-bath coupling) at late times. We then describe our open quantum system approach to inflation in section\ \ref{sec:systembath}, construct the modified evolution equation for the reduced density matrix in section\ \ref{sec:lindblad} and examine the time-dependence of the non-Hamiltonian terms in section \ \ref{sec:evaluate}. We conclude with a discussion in section\ \ref{sec:disc}. Various mathematical details are relegated to the appendices.


\section{The model}
\label{sec:model}

We work in a quasi-de Sitter space where the expansion is driven by a dynamically evolving scalar field. The background metric is $ds^2=-dt^2+a^2(t)d\vec{x}^2=-a^2(\eta)[d\eta^2-d\vec{x}^2]$, where $a$ is the scale factor, $t$ is cosmological time, and $-\infty\leq\eta\leq0$ is conformal time; dots (primes) indicate derivatives with respect to $t$ ($\eta$). Since the scalar field evolves, its energy density serves as a ``clock''  and provides a preferred choice of time slices. Each slice is spatially isotropic. In an expanding universe, physical wavelengths are stretched with time, and it is often convenient to work instead with comoving wavelengths, or comoving momenta $\vec{k}=a(\eta)\vec{p}$, $\vec{p}$ being the physical momentum, that remain invariant as a function of time.

The Hubble parameter $H=\dot{a}/a$ and its derivatives $\epsilon\equiv -\dot{H}/H^2$, $\delta\equiv \dot{\epsilon}/(H\epsilon)$, describe the time-evolution of the background. Quasi-de Sitter phases have $0<\epsilon<1$, so a nearly constant Hubble parameter. When $H$ is nearly constant we can integrate $a\,d\eta=dt$ to obtain the useful relation $\eta\approx -1/(aH)$. The quadratic action for the Fourier modes of the (dimensionless) scalar perturbation, $\zeta$, where $ds^2=-a^2(\eta)[d\eta^2-(1+2\zeta)d\vec{x}^2]$, is \cite{Mukhanov:1985rz,Sasaki:1986hm,Garriga:1999vw}
\bea
	S & = & \frac{1}{2}\int  \d \eta \int \frac{d^3 k}{(2\pi)^3} \, z^2 \( \zeta_{\vec{k}}' \zeta_{-\vec{k}}' - c_s^2k^2 \zeta_{\vec{k}} \zeta_{-\vec{k}} \) ,
\label{eq:zetaaction}
\eea
with $z^2=2\epsilon a^2 M_p^2/c_s^2$, where the (reduced) Planck mass $M_p$ is related to Newton's gravitational constant by $M_p^2=(8\pi G_N)^{-1}$ and $0<c_s\leq1$ is the sound speed.
To solve the evolution equations it is convenient to work with the canonical variable $\chi= z(\eta)\zeta$. Introducing creation and annihilation operators $\hat{c}$, $\hat{c}^{\dagger}$ that satisfy $\left[\hat{c}_{\vec{k}}(\eta),\hat{c}^{\dagger}_{\vec{k}^{\prime}}(\eta)\right]=(2\pi)^3\delta^3(\vec{k}-\vec{k}^{\prime})$ (and using the $\vec{k}\rightarrow-\vec{k}$ symmetry in the Fourier transform) we can write the Hamiltonian for the fluctuations \cite{Mukhanov:1981xt},
\bea
\label{eq:quadH}
	\hat{H} & = & \frac{1}{2} \int \frac{d^3 k}{(2\pi)^3} \bigg[ c_sk\( \hat{c}_{\vec{k}} \hat{c}_{\vec{k}}^{\dagger} + \hat{c}_{-\vec{k}} \hat{c}_{-\vec{k}}^{\dagger} \) \nonumber \\
	& & \quad - \ i \frac{z'}{z} \( \hat{c}_{\vec{k}} \hat{c}_{-\vec{k}} - \hat{c}_{\vec{k}}^{\dagger} \hat{c}_{-\vec{k}}^{\dagger} \) \bigg] .
\label{eq:hamladder}
\eea
This expression shows that the time-dependent gravitational background acts as a zero-momentum pump sourcing correlated pairs of $\chi$ quanta \cite{Grishchuk:1990cm, Albrecht:1992kf}. Notice that for a mode of fixed momentum $k$, the second line of the Hamiltonian is more important for $z^{\prime}/z\approx a^{\prime}/a=aH\gg c_sk$. In other words, the squeezing interaction term dominates the evolution when the physical wavelength of a mode is stretched to a scale larger than the Hubble size, $H^{-1}$. The broken time translation invariance ensures both that this scalar fluctuation cannot be gauged away, and gives an appropriate axis so that the two-mode squeezing introduced by the last term in the Hamiltonian is well-defined (i.e., the $\vec{k}$ and $-\vec{k}$ modes are distinguishable).


\section{Slow-roll versus non-attractor evolution}
\label{sec:srna}

In typical models of inflation (single-field, slow-roll), one chooses a slowly varying potential energy for the scalar field so that $\epsilon$ is nearly constant with a value typically of order 0.01 to 0.1. The equation of motion for the evolution of the field rapidly becomes independent of any initial velocity. The second-order differential equation of motion for $\zeta_{\vec{k}}$ gives rise to two different time-dependent pieces in the solution. These are commonly designated the ``growing" and ``decaying" modes, although in the case of standard slow-roll the ``growing" mode actually approaches a constant for $k\ll aH$ while the decaying mode rapidly becomes a negligible component of the solution. Then $\zeta_{\vec{k}}$ can be considered constant (and $\dot{\zeta}_{\vec{k}}\approx 0$) roughly from the time $k\lesssim aH\equiv -1/\eta_{\star}$ until the end of the inflationary phase, $\eta=0$. The observed curvature perturbation, $\zeta_{\vec{k}}$, and its conjugate momentum, $\pi_{-\vec{k}}$, satisfy $[\lim_{k\eta\rightarrow0^-}\zeta_{\vec{k}}(\eta),\lim_{k\eta\rightarrow0^-}\pi_{-\vec{k}}(\eta)]=0$, making quantum mechanical effects extremely difficult to observe even in the absence of any sources of decoherence. Since $\epsilon$ is nearly constant, the canonical field $\chi$ satisfies Eq.\ (\ref{eq:quadH}) with $z^{\prime}/z= a^{\prime}/a$ to a very good approximation.

A rather different behavior can be found if the potential for the scalar field has an exactly flat region, but the energy density is driven to evolve by giving the scalar field an initial velocity \cite{Kinney:2005vj}. Such a phase, often called a ``non-attractor phase" would only persist a short time, since the initial velocity is damped away by Hubble friction. But, while it lasts, it provides a background metric that is nearly de Sitter, but with $\epsilon\sim a^{-6}(\eta)$ far from constant and $\delta=-6$ not a small parameter. Assuming $c_s$ is a constant, $z^{\prime}/z=-2 a^{\prime}/a$. Crucially for our purposes, a change in dynamics affects the solution for $\zeta_{\vec{k}}$, which now has one contribution that grows as $\eta^{-3}$ and a second that is constant, so that $\dot{\zeta}_{\vec{k}}\neq$ constant even for $k\lesssim aH$. 

For either slow-roll or non-attractor dynamics the evolution of the canonical field $\chi$ at quadratic order can be found in terms of two-mode squeezing and rotation operators. One first needs to solve for the time-dependence of the ladder operators using the Heisenberg equation of motion,
\bea
	\frac{d \hat{c}_{\vec{k}}}{d\eta} & = & -i[\hat{c}_{\vec{k}}, \hat{H}] \nonumber \\
	& = & -i \( c_s k\hat{c}_{\vec{k}} + i\frac{z'}{z} \hat{c}_{-\vec{k}}^{\dagger} \) ,
\eea
which can in turn be solved by a Bogoliubov transformation with a choice of initial condition at time $\eta_0$, $\hat{c}_{\vec{k}}(\eta)=u_k(\eta) \hat{c}_{\vec{k}}(\eta_0)+v_k(\eta)\hat{c}_{-\vec{k}}^{\dagger}(\eta_0)$, where $|u_k(\eta)|^2-|v_k(\eta)|^2=1$. The Bogoliubov transformation can be written as
\begin{subequations}
\begin{align}
	u_k(\eta)&=e^{i\theta_k(\eta)}{\rm cosh}r_k(\eta) \, , \\
	v_k(\eta)&=e^{-i\theta_k(\eta)+2i\phi_k(\eta)}{\rm sinh}r_k(\eta) \, ,
\end{align}
\end{subequations}
where $r_k$ is the squeezing parameter, $\phi_k$ is the squeezing angle, and $\theta_k$ is an angle rotating the conjugate field and momenta (which is the same for the $\vec{k}$ and $-\vec{k}$ modes). The leading order time-dependence in the exact de Sitter background approximation for slow-roll inflation, is given by \cite{Albrecht:1992kf}
\begin{subequations}
\label{eq:rSR}
\begin{align}
r_k^{\rm SR}(\eta)&=-{\rm ArcSinh}\left(\frac{1}{2c_sk\eta}\right),\\
\phi_k^{\rm SR}(\eta)&=-\frac{\pi}{4}-\frac{1}{2}{\rm ArcTan}\left(\frac{1}{2c_sk\eta}\right),\\
\theta_k^{\rm SR}(\eta)&=-k\eta-{\rm ArcTan}\left(\frac{1}{2c_sk\eta}\right),
\end{align}
\end{subequations}
while for the non-attractor case we find instead
\begin{subequations}
\label{eq:rNA}
\begin{align}
r_k^{\rm NA}(\eta)&=-2\,{\rm ArcSinh}\left(\frac{3}{2c_sk\eta}\right),\\
\phi_k^{\rm NA}(\eta)&=\frac{\pi}{4}-\frac{1}{2}{\rm ArcTan}\left(\frac{3}{2c_sk\eta}\right),\\
\theta_k^{\rm NA}(\eta)&=-k\eta-2\sqrt{2}\;{\rm ArcTan}\left(\frac{3\sqrt{2}}{2c_sk\eta}\right).
\end{align}
\end{subequations}
For non-attractors the solution is approximate and only valid when $c_sk\eta\ll1$.

Note that the equation of motion for $\chi$ is the same in both, slow-roll and non-attractor models, i.e. $\chi_{\vec{k}}'' + \( c_s^2 k^2 - \frac{2}{\eta^2} \) \chi_{\vec{k}} = 0$. At any instant of time, however, the position $\chi_{\vec{k}}$ and conjugate momentum $p_{-\vec{k}}$ can be different between the two models, as indicated by the distinct time-dependence of the squeezing parameters above. The commutator, $[\chi_{\vec{k}}(\eta), p_{-\vec{k}'}(\eta)] = i (2\pi)^3 \delta^3(\vec{k} - \vec{k}')$, remains preserved at all times, as expected.


\section{Defining the system and bath}
\label{sec:systembath}

\begin{figure}[t!]
\centering
\includegraphics[width=.48\textwidth]{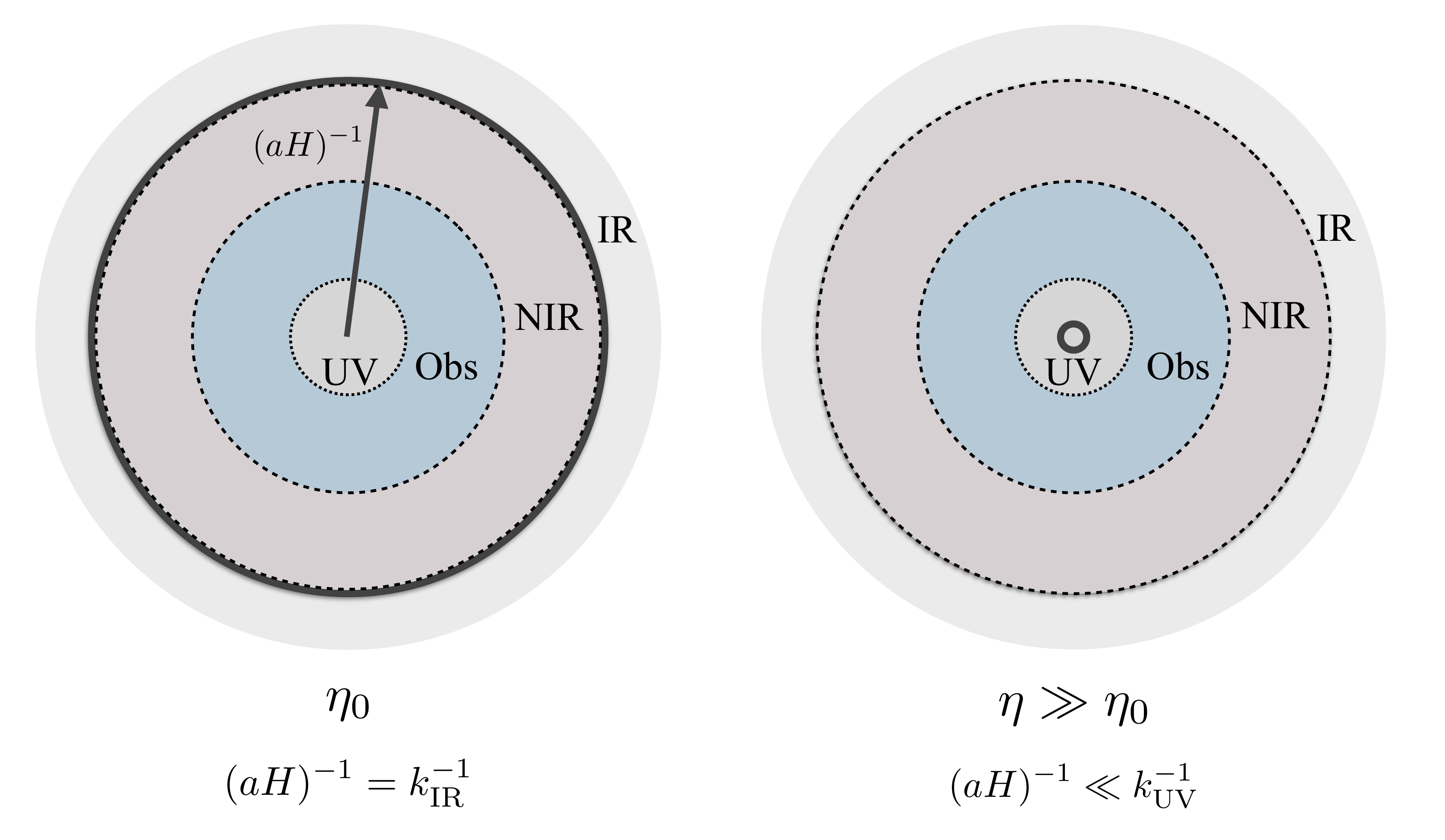}
\caption{\label{fig:sysbath} Representation of the system (``observable" modes) and bath (``near infrared", NIR, modes) Hilbert space, in terms of bands of comoving momenta. The comoving Hubble radius (thick black circle) is larger than all wavelengths of interest at the initial time $\eta_0$, but shrinks to be smaller than both bath and system wavelengths at late times. }
\end{figure}

We use bands of comoving momenta to define the system and bath, and assume that at some initial (conformal) time $\eta_0$ we can factorize the Hilbert space as\footnote{If we worked with physical momenta such a factorization should remain valid at all times, but, since we are using comoving momenta and the scale factor $a$ is fluctuating, the factorization would not quite hold. However, this prescription should capture the dominant features of the scenario (and we can check the physics using what is known about gauge issues for $\zeta_{\vec{k}}$ correlators).}
\be
\mathcal{H}=\mathcal{H}_{\rm UV}\otimes\mathcal{H}_{\rm Obs}\otimes\mathcal{H}_{\rm NIR}\otimes\mathcal{H}_{\rm IR} \, .
\ee
We focus only on how the evolution of modes in $\mathcal{H}_{\rm Obs}$, which satisfy $k_{\rm min}\leq k_{\rm Obs}\leq k_{\rm max}$, is affected by interactions with unobservable modes in the near infrared $\mathcal{H}_{\rm NIR}$, which satisfy $k_{\rm IR}<k_{\rm NIR}<k_{\rm min}$. We assume that modes in $\mathcal{H}_{\rm UV}$ can be properly accounted for with usual renormalization techniques. We also assume that modes far in the infrared, $k \in {\rm IR}$, were accounted for in defining the Hamiltonian at time $\eta_0$. This organization is shown diagrammatically in Fig.\ \ref{fig:sysbath}. Note that the comoving Hubble radius decreases as a function of time.

Further, we consider a cubic interaction term of the form, $S_3=M_p^2\int d^3x d\eta\, a^4(3\epsilon/c_s^2)(c_s^2-1)\zeta\dot{\zeta}^2$. Expressed in terms of the field $\chi$ in momentum space this reads
\ba
\lambda(\eta)\hat{H}_I&=&\frac{3(c_s^2-1)}{8M_pc_s^2a\sqrt{\epsilon}}\int_{\triangle}\left[\sqrt{\frac{k_2k_3}{k_1}}\left(\hat{c}^{\dagger}_{-\vec{k}_1}\hat{c}^{\dagger}_{-\vec{k}_2}\hat{c}^{\dagger}_{-\vec{k}_3}\right.\right.\nonumber\\
&&\hspace{0.5cm}\left.\left.+\,\hat{c}_{\vec{k}_1}\hat{c}^{\dagger}_{-\vec{k}_2}\hat{c}^{\dagger}_{-\vec{k}_3}+ \ldots \right)+{\rm perm.}\right] ,
\label{eq:cubicham}
\ea
where $\int_{\triangle}=\int\frac{d^3k_1}{(2\pi)^3}\frac{d^3k_2}{(2\pi)^3}\frac{d^3k_3}{(2\pi)^3}(2\pi)^3\delta^3(\vec{k}_1+\vec{k}_2+\vec{k}_3)$. The Dirac delta function enforces that the interacting momenta form a closed triangle, which is useful for categorizing contributions to the integral. The terms inside the parenthesis include all possible momentum conserving combinations of operators, with some terms appearing with a minus sign since the interaction term couples the field $\chi$ and its conjugate momentum. 

We choose this interaction term since it will significantly couple modes of different wavelengths in the non-attractor case where $\dot{\zeta}_{\vec{k}}$ does not become negligible on large scales \cite{Chen:2013eea}, but not in the slow-roll case. As with the quadratic Hamiltonian, the functional form is the same for the slow-roll and non-attractor cases; the difference is in the time-dependence of $\epsilon$. The coupling coefficient $\lambda(\eta)=3(c_s^2-1)/(8c_s^2a(\eta)\sqrt{\epsilon(\eta)})$ is dimensionless but time-dependent (we take $c_s$ to be constant for simplicity). Using the fact that $\epsilon(\eta)$ is approximately constant for slow-roll and $\sim a^{-6}(\eta)$ for non-attractor models, we obtain the following expressions for the coupling, with all time-dependence explicitly displayed:
\begin{subequations}
\begin{align}
\lambda^{\rm SR}(\eta)=&-\frac{3(c_s^2-1)}{8c_s^2\sqrt{\epsilon}}\eta H \, , \\
\lambda^{\rm NA}(\eta)=&\frac{3}{8}\frac{(c_s^2-1)}{c_s^2}\left(\frac{1}{H\eta}\right)^2.
\end{align}
\label{eq:lambdas}
\end{subequations}

Under the assumption that $\lambda(\eta)$ is abruptly turned on, and hence no system-bath coupling exists at $\eta_{0}$, the initial density matrix can be written as
\be
\hat{\sigma}(\eta_0)=|\psi_{\rm NIR}(\eta_0)\rangle|\psi_{\rm Obs}(\eta_0)\rangle\langle \psi_{\rm Obs}(\eta_0)|\langle \psi_{\rm NIR}(\eta_0)| \, .
\ee
The full time evolution is then given by $\hat{\sigma}(\eta)=\hat{U}(\eta,\eta_0)\hat{\sigma}(\eta_0)\hat{U}^{\dagger}(\eta,\eta_0)$, where the time evolution operator depends on the quadratic Hamiltonian for each mode, plus the relevant interaction term ($\hat{\sigma}(\eta)$ is the Schr\"{o}dinger picture density matrix). We note that both the quadratic Hamiltonian, containing the two-mode squeezing term, and the cubic interaction are time-dependent. However, for small initial coupling the full evolution can be approximated as
\ba
\hat{U}(\eta,\eta_0)&=&Te^{-i\int_{\eta_0}^{\eta}\hat{H}_0(\eta_1)d\eta_1}Te^{-i\int_{\eta_0}^{\eta} \hat{H}_{I,i}(\eta_1)d\eta_1},
\label{eq:U(t)}%
\ea
where $T$ time-orders the exponentials and $\hat{H}_{I,i}$ is the interaction Hamiltonian in the interaction picture.

To perform the trace over the near infrared degrees of freedom, we introduce two kinds of basis states for the bath modes:
\begin{itemize}
\itemsep=-2pt
\item Fock states defined at $\eta_0$ grouped into $(\vec{k}$, $-\vec{k})$ pairs, as $|N\rangle=\prod_{k\in {\rm NIR}}|m_{\vec{k}},n_{-\vec{k}}\rangle$. Summing over $|N\rangle$ amounts to summing over all possible pairs of integer values for $m_{\vec{k}}$ and $n_{-\vec{k}}$. These are eigenstates of the quadratic Hamiltonian {\it without} the squeezing term. 
\item The two-mode squeezed vacuum for the bath modes, represented by the action of the propagator, corresponding to the full quadratic Hamiltonian for the bath, on the vacuum: $\hat{U}_0(\eta,\eta_0)|0_{\vec{k}},0_{-\vec{k}}\rangle \equiv|SQ(k,\eta)\rangle=\sum_n c^{\rm sq}_n(k,\eta)|n_{\vec{k}},n_{-\vec{k}}\rangle$. Note that, unlike $|N\rangle$, the squeezed vacuum is explicitly time-dependent due to the time-dependence of $(r_{k}, \phi_{k}, \theta_{k})$ in Eqs.\ (\ref{eq:rSR}) and (\ref{eq:rNA}). 
\end{itemize}


\section{The evolution equation}
\label{sec:lindblad}

The reduced density matrix for the observable modes, at any time $\eta\geq\eta_0$, is given by
\ba
\label{eq:rho(eta)}
\hat{\rho}(\eta)&=&{\rm Tr}_{\rm NIR}\;\hat{\sigma}(\eta)\nonumber\\
 &=&\sum_{N}\langle N|\hat{U}(\eta,\eta_0)|\psi_{\rm NIR}(\eta_0)\rangle|\psi_{\rm Obs}(\eta_0)\rangle\nonumber\\
 &&\hspace{0.65cm}\times\,\langle \psi_{\rm Obs}(\eta_0)|\langle \psi_{\rm NIR}(\eta_0)|\hat{U}^{\dagger}(\eta,\eta_0)|N\rangle \, , \quad
\ea
where $\hat{U}(\eta,\eta_0)$ is given by Eq.\ (\ref{eq:U(t)}). Perturbatively expanding the above equation to second order in the coupling, we find that (see appendix\ \ref{app:evol} for details)
\begin{align}
\partial_{\eta}\hat{\rho}(\eta)=&-i\left[\hat{H}^{\rm Obs}_{0},\hat{\rho}(\eta)\right]-i\left[\hat{H}_{\rm eff},\hat{\rho}^{(0)}(\eta)\right]\nonumber\\
& +\,\{\hat{A}(\eta),\hat{\rho}^{(0)}(\eta)\}\nonumber\\
& +\sum_N\left[ \hat{L}_{N1}\hat{\rho}^{(0)}(\eta)\hat{L}_{N2}^{\dagger} + \hat{L}_{N2}\hat{\rho}^{(0)}(\eta)\hat{L}_{N1}^{\dagger}\right] ,
\label{eq:Inf_results}%
\end{align}
with\footnote{It would be interesting to consider the connection between the effective evolution derived in this way to the well-studied case of warm inflation \cite{Berera:2008ar,Bastero-Gil:2016qru}, where extra degrees of freedom lead to significant dissipative terms in the classical Lagrangian for the inflaton.} $\hat{H}_{\rm eff}^{(1)}=\lambda(\eta)\langle SQ(\eta)|\hat{H}_I(\eta_0)|SQ(\eta)\rangle$,
\begin{subequations}
\begin{align}
\hat{H}_{\rm eff}^{(2)}&=-\frac{i}{2}\sum_N(\hat{L}_{N1}^{\dagger}\hat{L}_{N2}-\hat{L}_{N2}^{\dagger}\hat{L}_{N1}) \, ,\\
\hat{A}(\eta)&=-\frac{1}{2}\sum_N \left(\hat{L}_{N1}^{\dagger}\hat{L}_{N2}+\hat{L}_{N2}^{\dagger}\hat{L}_{N1}\right) ,
\end{align}
\label{eq:HeffA}%
\end{subequations}
and the Lindblad operators given by
\begin{subequations}
\begin{align}
\hat{L}_{N1}(\eta)&=\lambda(\eta)\langle N|\hat{H}_I(\eta_0)|SQ(\eta)\rangle \, ,\\
\hat{L}_{N2}(\eta)&=\int_{\eta_0}^{\eta} d\eta_1\lambda(\eta_1)\langle N|\hat{H}_{I,i}(\eta_1-\eta)|SQ(\eta)\rangle.
\end{align}
\label{eq:LNs}%
\end{subequations}
Here $|SQ(\eta)\rangle=\prod_{k\in{\rm NIR}}|SQ(k,\eta)\rangle$ and $\hat{H}_0^{\rm Obs}$ is defined by restricting the integral in Eq.\ (\ref{eq:quadH}) to only run over modes $k\in {\rm Obs}$. This result in Eqs. (\ref{eq:Inf_results})-(\ref{eq:LNs}) is similar to that of \cite{Agon:2014uxa}, but with additional structure due to the time-dependent squeezing term at quadratic order.

\section{Evaluating the non-Hamiltonian evolution}
\label{sec:evaluate}
The separation between system and bath is in momentum space, so we must work there to find explicit expressions for the non-Hamiltonian terms in the evolution of $\hat{\rho}(\eta)$. 
As the first check on the expressions above, suppose all three momenta are in the NIR bath. Then the $\hat{L}_{Ni}$ are just numbers and so $\hat{H}_{\rm eff}^{(2)}=0$ and the terms in the last two lines of Eq.\ (\ref{eq:Inf_results}) all sum to zero. For momentum configurations containing both system and bath modes, the fact that all non-Hamiltonian terms in the evolution equation come with $\sum_N$ ensures they will give non-zero contributions only when the same number of modes are in the NIR in both $\hat{L}_{Ni}$ and $\hat{L}^{\dagger}_{Nj}$. That, in turn, means all terms in the evolution equation contain an even number of $\hat{c}_{\vec{k}}$, $\hat{c}^{\dagger}_{\vec{k}}$ operators for modes in the observable band. 

The momentum configurations that give non-zero non-Hamiltonian evolution can be conveniently thought of in the language familiar from the study of non-Gaussianity in cosmology. They are either (1) ``folded" triangles, where two bath modes interact with one system mode or (2) ``squeezed'' triangles, where two system modes interact with one bath mode. Only system modes with momenta $k_{\rm min}<k<2k_{\rm min}$ can receive contributions of the folded type, and even for these selected modes not many configurations are possible. The fact that the same bath state $|N\rangle$ appears in both Lindblad operators in terms like $\hat{L}_{N1}\hat{\rho}^{(0)}(\eta)\hat{L}_{N2}^{\dagger}$, enforces conservation of momentum of the system modes appearing explicitly in the final result. 
\par
In appendix\ \ref{app:lind}, we write out the interaction Hamiltonian and the two Lindblad operators for folded and squeezed configurations. After specifying the bath modes for each case, we evaluate all creation, annihilation, or squeezing operators acting on bath modes. From the results for folded and squeezed configurations (Eqs.\ (\ref{eq:LN1fold}), (\ref{eq:LN2fold}) and Eqs.\ (\ref{eq:LN1sq}), (\ref{eq:LN2sq}) respectively), the sum $\sum_N\hat{L}_{N1}\hat{\rho}^{(0)}(\eta)\hat{L}_{N2}^{\dagger}$ can be evaluated. 

\begin{figure*}[t!]
\centering
\subfigure[
]{\includegraphics[width=0.475\textwidth]{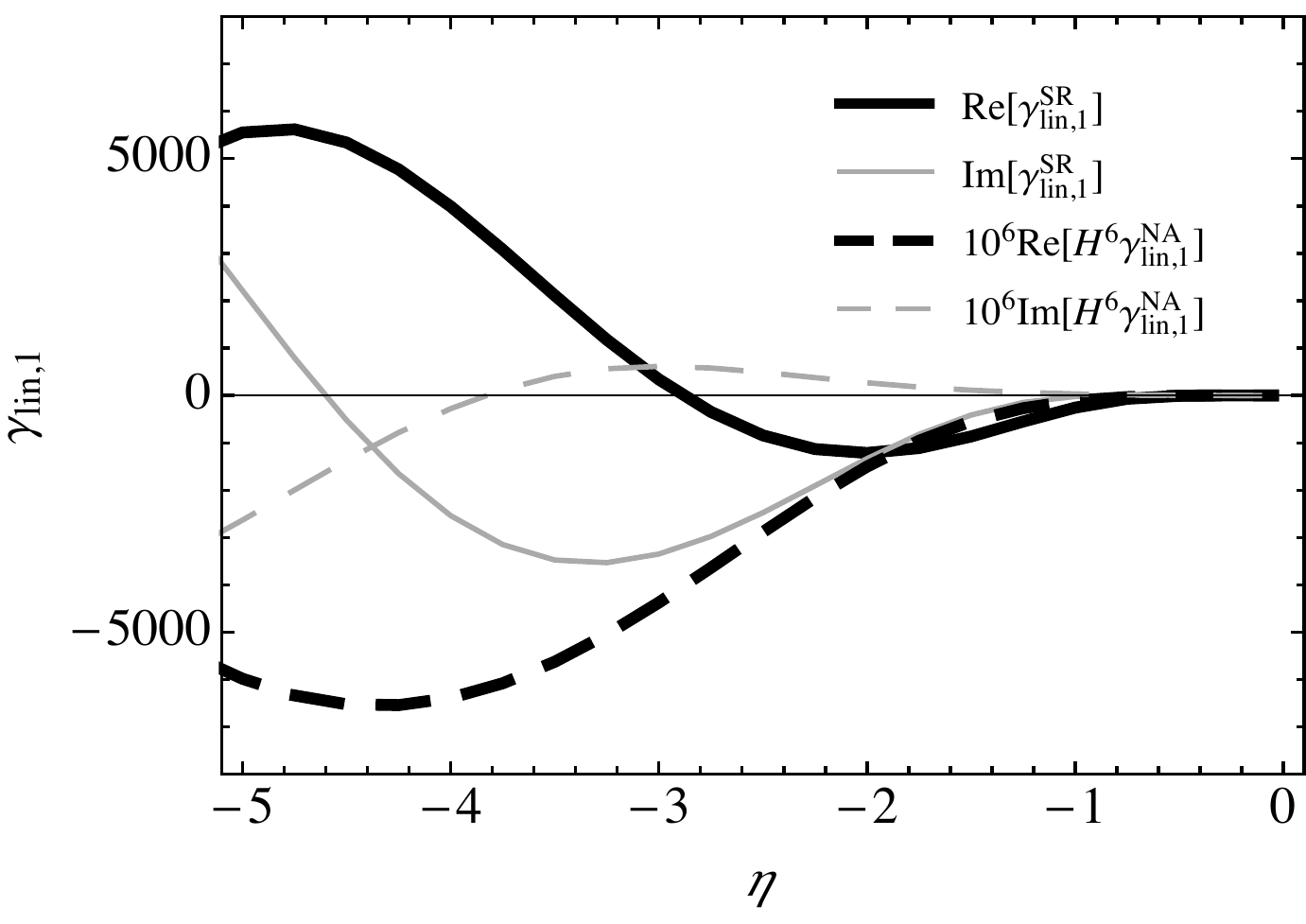}}
\subfigure[
]{\includegraphics[width=0.475\textwidth]{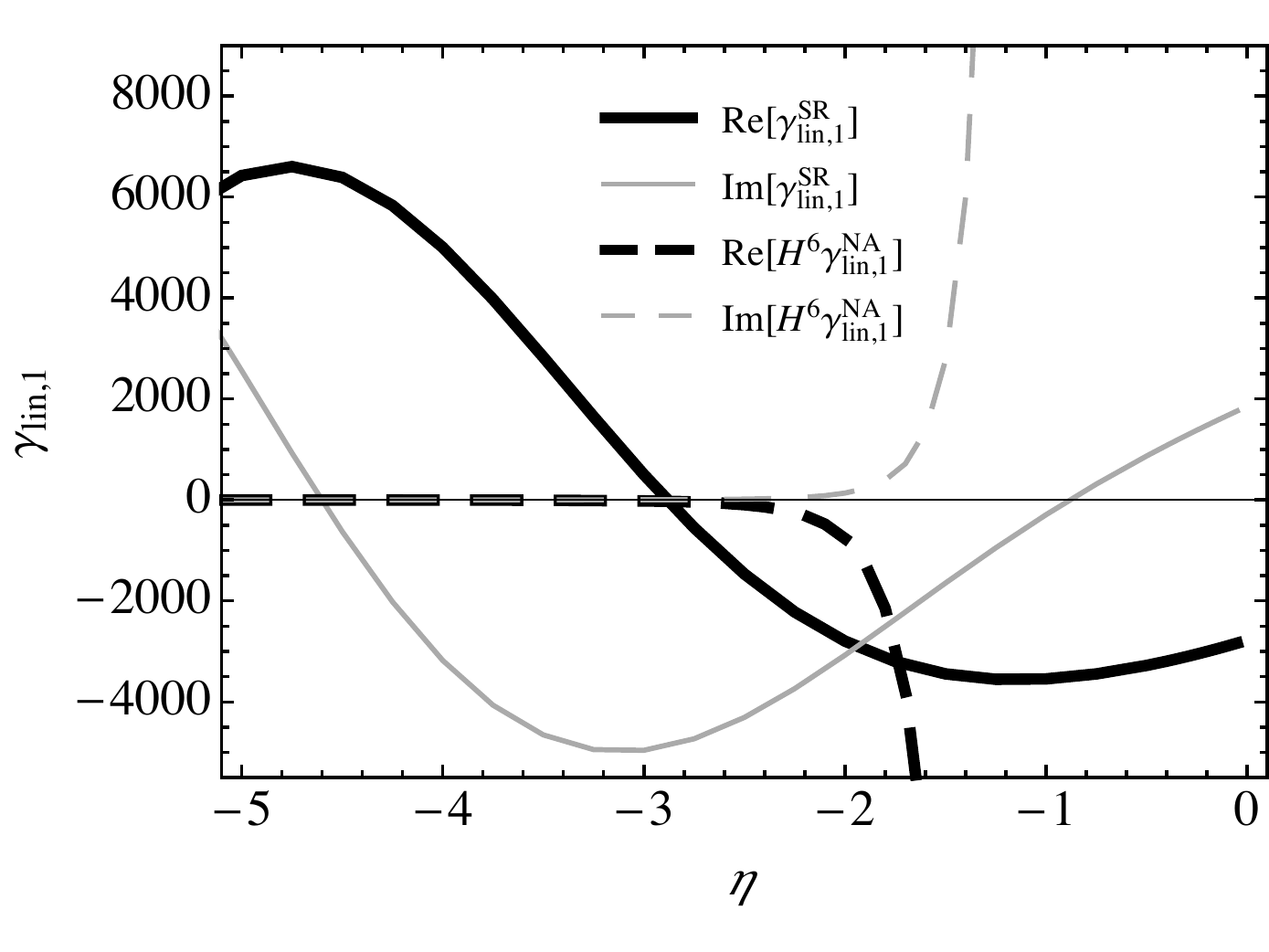}}
\caption{Example contribution to $\sum_N\hat{L}_{N1}(\eta) \hat{L}_{N2}^{\dagger}(\eta)$ from a folded triangle configuration with momenta (in units of $(H/c_s)$) of $k_s=1$, $k_{b1}=0.5$ and $k_{b2}=0.54$ and bath modes in (a) the quantum ground state, i.e. $m_{\vec{k}_{b1}} = m_{\vec{k}_{b2}} = 0$, and (b) an arbitrary superposition of Fock states, i.e. summing over all possible values of $m_{\vec{k}_{b1}}$ and $m_{\vec{k}_{b2}}$. For both slow-roll (SR) and non-attractor (NA) dynamics we extract the dimensionless parameter $\gamma_{{\rm lin},1}$ from $f_1$ in Eq.\ (\ref{eq:DissFolded}), but the non-attractor case is shown in units of $H^{-6}$ so as not to obscure the dependence of the amplitude on this physical number. As seen from Eq.\ (\ref{eq:HeffA}), both the real and imaginary parts of this quantity enter the evolution equation.}
\label{fig:Gammaslin}
\end{figure*}

For example, consider a folded triangle with labels $\vec{k}_{s}$, $\vec{k}_{b1}$, $\vec{k}_{b2}$ for the momenta in $\hat{L}_{N1}$ and $\vec{k}^{\prime}_{s}$, $\vec{k}^{\prime}_{b1}$, $\vec{k}^{\prime}_{b2}$ in $\hat{L}^{\dagger}_{N2}$. Then,
\begin{widetext}
\begin{align}
&\left.\sum_N\,\hat{L}_{N1}\hat{\rho}^{(0)}(\eta)\hat{L}_{N2}^{\dagger}\right|_{\rm Folded}\nonumber \\
&\hspace{1cm}=\sum_{m_{\vec{k}_{bi}}, n_{\vec{k}_{bi}}}\frac{\lambda(\eta)}{M_p^2}\int_{\triangle}\int_{\triangle^{\prime}}\prod_{k_i\in {\rm NIR}, k_i\neq k_{b1}, k_{b2}}\langle m_{\vec{k}_i},n_{-\vec{k}_i}|SQ(k_i,\eta)\rangle\prod_{k^{\prime}_i\in {\rm NIR}, k^{\prime}_i\neq k^{\prime}_{b1}, k^{\prime}_{b2}}\langle SQ(k_i^{\prime},\eta)| m_{\vec{k}^{\prime}_i},n_{-\vec{k}^{\prime}_i}\rangle\nonumber\\
&\hspace{1.5cm}\times\frac{e^{+i\theta_{k^{\prime}_{b1}}(\eta)}}{{\rm cosh}\,r_{k^{\prime}_{b1}}(\eta)}\frac{e^{+i\theta_{k^{\prime}_{b2}}(\eta)}}{{\rm cosh}\,r_{k^{\prime}_{b2}}(\eta)}\;c^{\rm sq\,*}_{m_{\vec{k}^{\prime}_{b1}}}(k^{\prime}_{b1},\eta)c^{\rm sq\,*}_{m_{\vec{k}^{\prime}_{b2}}}(k^{\prime}_{b2},\eta)\frac{1}{(k_{b1}k_{b2})^{3/2}}\frac{1}{(k^{\prime}_{b1}k^{\prime}_{b2})^{3/2}}
\nonumber\\
&\hspace{1.5cm}\times \delta_{m_{\vec{k}^{\prime}_{b1}}+1,n_{\vec{k}^{\prime}_{b1}}}\delta_{m_{\vec{k}^{\prime}_{b2}}+1,n_{\vec{k}^{\prime}_{b2}}}\delta_{m_{\vec{k}_{b1}}+1,n_{\vec{k}_{b1}}}\delta_{m_{\vec{k}_{b2}}+1,n_{\vec{k}_{b2}}}\sqrt{(m_{\vec{k}^{\prime}_{b1}}+1)(m_{\vec{k}^{\prime}_{b2}}+1)}\sqrt{(m_{\vec{k}_{b1}}+1)(m_{\vec{k}_{b2}}+1)}\nonumber\\
&\hspace{1.5cm}\times\left\{\hat{c}_{\vec{k}_s}(\eta_0)\left[\sqrt{\frac{k_{b1}k_{b2}}{k_s}}\left[c^{\rm sq}_{n_{\vec{k}_{b1}}}(k_{b1},\eta)c^{\rm sq}_{n_{\vec{k}_{b2}}}(k_{b2},\eta)+c^{\rm sq}_{m_{\vec{k}_{b1}}}c^{\rm sq}_{m_{\vec{k}_{b2}}}-c^{\rm sq}_{n_{\vec{k}_{b1}}}c^{\rm sq}_{m_{\vec{k}_{b2}}}-c^{\rm sq}_{m_{\vec{k}_{b1}}}c^{\rm sq}_{n_{\vec{k}_{b2}}}\right]+\dots\right]\right.\nonumber\\
&\hspace{1.8cm}+\left.\hat{c}^{\dagger}_{-\vec{k}_s}(\eta_0)\left[\sqrt{\frac{k_{b1}k_{b2}}{k_s}}[++--]+\sqrt{\frac{k_sk_{b2}}{k_{b1}}}\,[-++-]+\sqrt{\frac{k_sk_{b1}}{k_{b2}}}\,[-+-+]\right]\right\}\,\hat{\rho}^{(0)}(\eta)\hat{S}_{k^{\prime}_{s}}(\eta)\hat{R}_{k^{\prime}_{s}}(\eta)\nonumber\\
&\hspace{1.5cm}\times\left\{\hat{c}^{\dagger}_{\vec{k}^{\prime}_{s}}(\eta_0)\hat{R}^{\dagger}_{k^{\prime}_{s}}(\eta)\hat{S}^{\dagger}_{k^{\prime}_{s}}(\eta)\int_{\eta_0}^{\eta}d\eta_1\lambda(\eta_1)\left\{\sqrt{\frac{k^{\prime}_{b1}k^{\prime}_{b2}}{k^{\prime}_{s}}}\left[u^*_{k^{\prime}_s}(\eta_1)\,\left(v^*_{k^{\prime}_{b1}}(\eta_1)v^*_{k^{\prime}_{b2}}(\eta_1)+uu-v^*u-uv^*\right)\right.\right.\right.\nonumber\\
&\hspace{8.1cm}\left.\left.+ \, v(k^{\prime}_{\bm s},\eta_1)(++--)\right]+\dots\right\}\nonumber\\
&\hspace{1.8cm}\left.+\,\hat{c}_{-\vec{k}^{\prime}_{s}}(\eta_0)\hat{R}^{\dagger}_{k^{\prime}_{s}}(\eta)\hat{S}^{\dagger}_{k^{\prime}_{s}}(\eta)\int_{\eta_0}^{\eta}d\eta_1\lambda(\eta_1)\left\{\sqrt{\frac{k^{\prime}_{b1}k^{\prime}_{b2}}{k^{\prime}_{s}}}\left[v^*_{k^{\prime}_s}(\eta_1)(++--)+u_{k^{\prime}_s}(\eta_1)(++--)\right]+\dots\,\right\}\right\} ,
\end{align}
\end{widetext}
where subscripts `s'  denote system modes, subscripts `b' denote bath modes, and $\hat{S}_{k}(\eta)$ and $\hat{R}_{k}(\eta)$ are the two-mode squeezing and rotation operators constructed from $(r_{k}, \phi_{k}, \theta_{k})$ in Eqs.\ (\ref{eq:rSR}) and (\ref{eq:rNA}). Since the same bath state $|N\rangle$ appears in both Lindblad operators (Eq.\ (\ref{eq:LNs})), the momenta appearing in either operator must also be the same (e.g., $\vec{k}_{b1}^{\prime}= \vec{k}_{b1}$, where the Dirac deltas for the triangle modes enforce the relative sign in this equality). Using this fact (and replacing the integrals by symmetrized products $(1/3)\sum_{i,j} k_{i}^{3}k_{j}^3$ to maintain the correct dimensionality) gives
\begin{widetext}
\begin{align}
&\left.\sum_N\,\hat{L}_{N1}\hat{\rho}^{(0)}(\eta)\hat{L}_{N2}^{\dagger}\right|_{\rm Folded}\nonumber\\
&\hspace{1cm}=\sum_{m_{\vec{k}_{bi}}, n_{\vec{k}_{bi}}}\frac{\lambda(\eta)}{M_p^2}\int_{\triangle}\prod_{k_i\in {\rm NIR}, k_i\neq k_{b1}, k_{b2}}|\langle m_{\vec{k}_i},n_{-\vec{k}_i}|SQ(k_i,\eta)\rangle|^2\,\delta_{m_{\vec{k}_{b1}}+1,n_{\vec{k}_{b1}}}\delta_{m_{\vec{k}_{b2}}+1,n_{\vec{k}_{b2}}}\left[\frac{k_{b1}^3k_{b2}^3+k_s^3k_{b1}^3+k_s^3k_{b2}^3}{3k_{b1}^3k_{b2}^3}\right]\nonumber\\
&\hspace{1.5cm}\times\frac{e^{+i\theta_{k_{b1}}(\eta)}}{{\rm cosh}\,r_{k_{b1}}(\eta)}\frac{e^{+i\theta_{k_{b2}}(\eta)}}{{\rm cosh}\,r_{k_{b2}}(\eta)}(m_{\vec{k}_{b1}}+1)(m_{\vec{k}_{b2}}+1)\;c^{\rm sq\,*}_{m_{\vec{k}_{b1}}}(k_{b1},\eta)c^{\rm sq\,*}_{m_{\vec{k}_{b2}}}(k_{b2},\eta)\nonumber\\
&\hspace{1.5cm}\times\left\{\hat{c}_{\vec{k}_{s}}(\eta_0)\left[\sqrt{\frac{k_{b1}k_{b2}}{k_s}}\left[c^{\rm sq}_{n_{\vec{k}_{b1}}}(k_{b1},\eta)c^{\rm sq}_{n_{\vec{k}_{b2}}}(k_{b2},\eta)+c^{\rm sq}_{m_{\vec{k}_{b1}}}c^{\rm sq}_{m_{\vec{k}_{b2}}}-c^{\rm sq}_{n_{\vec{k}_{b1}}}c^{\rm sq}_{m_{\vec{k}_{b2}}}-c^{\rm sq}_{m_{\vec{k}_{b1}}}c^{\rm sq}_{n_{\vec{k}_{b2}}}\right]+\dots\right]\right.\nonumber\\
&\hspace{1.8cm}+\left.\hat{c}^{\dagger}_{-\vec{k}_{s}}(\eta_0)\left[\sqrt{\frac{k_{b1}k_{b2}}{k_s}}[++--]+\sqrt{\frac{k_sk_{b2}}{k_{b1}}}\,[-++-]+\sqrt{\frac{k_sk_{b1}}{k_{b2}}}\,[-+-+]\right]\right\}\,\hat{\rho}^{(0)}(\eta)\hat{S}_{k_s}(\eta)\hat{R}_{k_s}(\eta)\nonumber\\
&\hspace{1.5cm}\times\left\{\hat{c}^{\dagger}_{\vec{k}_{s}}(\eta_0)\hat{R}^{\dagger}_{k_s}(\eta)\hat{S}^{\dagger}_{k_s}(\eta)\int_{\eta_0}^{\eta}d\eta_1\lambda(\eta_1)\left\{\sqrt{\frac{k_{b1}k_{b2}}{k_{s}}}\left[u^*_{k_s}(\eta_1)\,\left(v^*_{k_{b1}}(\eta_1)v^*_{k_{b2}}(\eta_1)+uu-v^*u-uv^*\right)\right.\right.\right.\nonumber\\
&\hspace{8.1cm}\left.\left.+\,v_{k_s}(\eta_1)(++--)\right]+\dots\right\}\nonumber\\
&\hspace{1.8cm}\left.+\,\hat{c}_{-\vec{k}_{s}}(\eta_0)\hat{R}^{\dagger}_{k_s}(\eta)\hat{S}^{\dagger}_{k_s}(\eta)\int_{\eta_0}^{\eta}d\eta_1\lambda(\eta_1)\left\{\sqrt{\frac{k_{b1}k_{b2}}{k_{s}}}\left[v^*_{k_s}(\eta_1)(++--)+u_{k_s}(\eta_1)(++--)\right]+\dots\,\right\}\right\} .
\end{align}
\end{widetext}
This result contains a sum over the bath modes participating in the interaction, whose structure depends on the fact that the bath is squeezed. (To illustrate how this bath structure is relevant, the left hand panel of Fig.\ \ref{fig:Gammaslin} below will contrast the actual dissipation compared to what results from considering just the ground state of the bath.) In the various terms of the interaction Hamiltonian, there are several versions of the mode sum that must be performed, but they all have the same form. For example, the third line of the right hand side in the previous equation (making use of the Dirac deltas from the first line) can be simplified using
\begin{align}
&\sum_{m_{\vec{k}_{b1}}=0}^{\infty}\sum_{m_{\vec{k}_{b2}}=0}^{\infty}(m_{\vec{k}_{b1}}+1)(m_{\vec{k}_{b2}}+1)\nonumber\\
&\hspace{0.2cm}\times c^{\rm sq\,*}_{m_{\vec{k}_{b1}}}(k_{b1},\eta)c^{\rm sq\,*}_{m_{\vec{k}_{b2}}}(k_{b2},\eta) \nonumber \\
&\hspace{0.2cm}\times\left[c^{\rm sq}_{m_{\vec{k}_{b1}}+1}(k_{b1},\eta)c^{\rm sq}_{m_{\vec{k}_{b2}}+1}(k_{b2},\eta)+c^{\rm sq}_{m_{\vec{k}_{b1}}}c^{\rm sq}_{m_{\vec{k}_{b2}}}\right.\nonumber\\
&\left.\hspace{1cm}-\ c^{\rm sq}_{m_{\vec{k}_{b1}}+1}c^{\rm sq}_{m_{\vec{k}_{b2}}}-c^{\rm sq}_{m_{\vec{k}_{b1}}}c^{\rm sq}_{m_{\vec{k}_{b2}}+1}\right] \nonumber \\
=&\,{\rm cosh}^3r_{k_{b1}}{\rm cosh}^3r_{k_{b2}}\nonumber\\
&\hspace{0.2cm}\times\left[c_0^{\rm sq}(k_{b1},\eta)-c_1^{\rm sq}(k_{b1},\eta)\right][c_0^{\rm sq}(k_{b2},\eta)-c_1^{\rm sq}(k_{b2},\eta)]
\end{align} 
The sums in other terms give similar results, but with varying signs in front of the $c_0^{\rm sq}$,  $c_1^{\rm sq}$ terms. 

Thus, folded triangles lead to terms in $\sum_N\,\hat{L}_{N1}\hat{\rho}^{(0)}(\eta)\hat{L}_{N2}^{\dagger}$ that, in terms of operators for system modes only, are of the form
\begin{align}
&\int\frac{d^3k_s}{(2\pi)^3}\hat{c}_{\vec{k}_{s}}(\eta_0)\hat{\rho}^{(0)}(\eta)\hat{S}_{k_s}(\eta)\hat{R}_{k_s}(\eta)\nonumber\\
&\hspace{0.5cm}\times\left[\hat{c}^{\dagger}_{\vec{k}_s}(\eta_0)f_1(k_i,\eta)+\hat{c}_{-\vec{k}_s}(\eta_0)f_2(,k_i,\eta)\right]\nonumber\\
&\hspace{0.5cm}\times\hat{R}^{\dagger}_{k_s}(\eta)\hat{S}^{\dagger}_{k_s}(\eta) \\ \nonumber
&{\rm (plus \; two \; more \; similar \; terms)}
\end{align}
where the functions $f_i$ have mass dimension 1. To define a dimensionless quantity $\gamma_{{\rm lin},i}$, signifying a ``dissipation factor"  for each momentum configuration, the integral over the triangle configuration can be simplified, leaving
\begin{align}
\label{eq:gam}
f^{\rm SR}_i \, \equiv \, &\,2\pi\frac{H^2}{M_p^2}\int_{k_{\rm IR}}^{k_{\rm min}}dk_b \int_{-1}^1du\, \theta_H(k_{\rm min}^2-\tilde{k}_b^2) \nonumber\\
&\times\left[\left(\frac{3(c_s^2-1)}{8c_s^2\sqrt{\epsilon}}\right)\right]^2 \gamma^{\rm SR}_{{\rm lin},i} \, ,\nonumber\\
f^{\rm NA}_i \, \equiv \, &2\pi\frac{H^2}{M_p^2}\int_{k_{\rm IR}}^{k_{\rm min}}dk_b \int_{-1}^1du\, \theta_H(k_{min}^2-\tilde{k}^2)\nonumber\\
&\times \left[\frac{3(c_s^2-1)}{8c_s^2}\right]^2 \gamma^{\rm NA}_{{\rm lin},i}  \, .
\end{align}
Repeating the analysis for squeezed configurations leads to dimensionless nonlinear dissipation factors defined by
\begin{align}
	q_{i}^{\rm SR} \, = \, & 2\pi \frac{H^{2}}{M_{p}^{2}}\int_{-1}^{1}du\;\theta_H(k_{\rm min}^2-k_b^2)\nonumber\\
&\times \left[-\left(\frac{3(c_s^2-1)}{8c_s^2 \sqrt{\epsilon}}\right)\right]^2\sqrt{k_{s1}k_{s2}}\gamma_{{\rm NL},i}^{\rm SR} \, , \nonumber\\
q_{i}^{\rm NA} \, = \, &2\pi \frac{H^{2}}{M_{p}^{2}}\int_{-1}^{1}du\;\theta_H(k_{\rm min}^2-k_b^2)\nonumber\\
 &\times\left[-\left(\frac{3(c_s^2-1)}{8c_s^2}\right)\right]^2\sqrt{k_{s1}k_{s2}}\gamma_{{\rm NL},i}^{\rm NA} \, .
\end{align}

Then, the general result for the form of terms in $\sum_N\hat{L}_{N1}\hat{\rho}^{(0)}(\eta)\hat{L}_{N2}^{\dagger}$ originating from a cubic interaction between system and bath modes is
\begin{widetext}
\begin{eqnarray}
\sum_N\hat{L}_{N1}\hat{\rho}^{(0)}(\eta)\hat{L}_{N2}^{\dagger}\Big|_{\rm Folded} &\supset &\int\frac{d^3k_s}{(2\pi)^3}\hat{c}_{\vec{k}_s}(\eta_0)\hat{\rho}^{(0)}(\eta)\hat{S}_{k_s}(\eta)\hat{R}_{k_s}(\eta) \nonumber\\
& & \qquad \qquad \times \left[\hat{c}^{\dagger}_{\vec{k}_s}(\eta_0)f_1(k_{i},\eta)+\hat{c}_{-\vec{k}_s}(\eta_0)f_2(k_{i},\eta)\right]\hat{R}^{\dagger}_{k_s}(\eta)\hat{S}^{\dagger}_{k_s}(\eta) \, , \qquad
\label{eq:DissFolded}
\\
\sum_N \hat{L}_{N1}\hat{\rho}^{(0)}(\eta)\hat{L}_{N2}^{\dagger}\Big|_{\rm Squeezed} &\supset & \int\frac{d^3k_{s1}}{(2\pi)^3}\int dk_{s2} \; k_{s2}^{2}\hat{c}_{\vec{k}_{s1}}(\eta_0)\hat{c}_{\vec{k}_{s2}}(\eta_0)\hat{\rho}^{(0)}(\eta)
\hat{S}_{k_{s1}}(\eta)\hat{R}_{k_{s1}}(\eta)\hat{S}_{k_{s2}}(\eta)\hat{R}_{k_{s2}}(\eta)\nonumber\\
& & \qquad \qquad \times \left[\hat{c}^{\dagger}_{\vec{k}_{s1}}(\eta_0)\hat{c}^{\dagger}_{\vec{k}_{s2}}(\eta_0)q_1(k_i,\eta) 
+ \dots \right]
\hat{R}^{\dagger}_{k_{s2}}(\eta)\hat{S}^{\dagger}_{k_{s2}}(\eta)\hat{R}^{\dagger}_{k_{s1}}(\eta)\hat{S}^{\dagger}_{k_{s1}}(\eta) \, ,
\label{eq:DissSq}
\end{eqnarray}
\end{widetext}
where ($\dots$) denotes all possible momentum-conserving operator pairs. Besides the time-dependence of the dissipation factors coming from the interaction Hamiltonian and time-dependent squeezing of the bath modes, the non-Hamiltonian evolution terms for the density matrix of the system also have time-dependence from the squeezing of the system modes (the $\hat{R}_{k_{s}}(\eta)$, $\hat{S}_{k_{s}}(\eta)$ operators in the equations above).

As evident from the explicit structure of the non-Hamiltonian terms, folded configurations lead to linear (single-mode) dissipation terms, while squeezed configurations lead to nonlinear (two-mode) dissipation terms. Further in each of the configurations, there are two classes of terms: $\big(\hat{C}_{k} \hat{\rho} \hat{C}_{k}^{\dagger}, \, \hat{C}_{k}^{\dagger} \hat{\rho} \hat{C}_{k}\big)$ and $\big( \hat{C}_{k} \hat{\rho} \hat{C}_{k}, \, \hat{C}_{k}^{\dagger} \hat{\rho} \hat{C}_{k}^{\dagger}\big)$ (here $\hat{C}_{k} \equiv \hat{c}_{k_{s}}$ for folded and $\hat{C}_{k} \equiv \hat{c}_{k_{s1}}\hat{c}_{k_{s2}}$ for squeezed configurations). While the former correspond to single/two-photon exchange with thermally distributed  bath modes, the latter terms indicate that these bath modes are squeezed \cite{Scully:1997}. 

\begin{figure*}[t!]
\centering
\subfigure[
]{\includegraphics[width=0.45\textwidth]{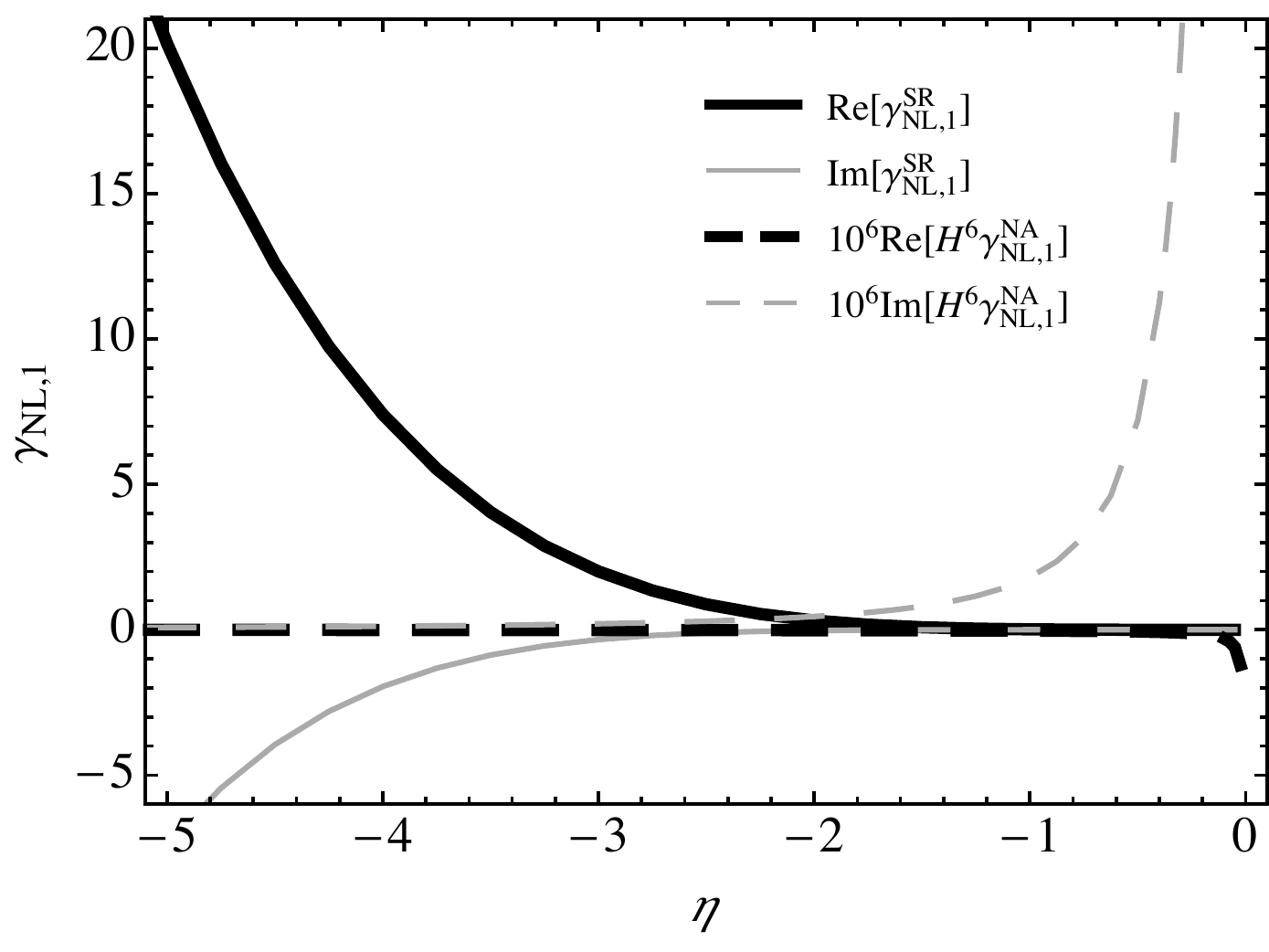}}
\subfigure[
]{\includegraphics[width=0.49\textwidth]{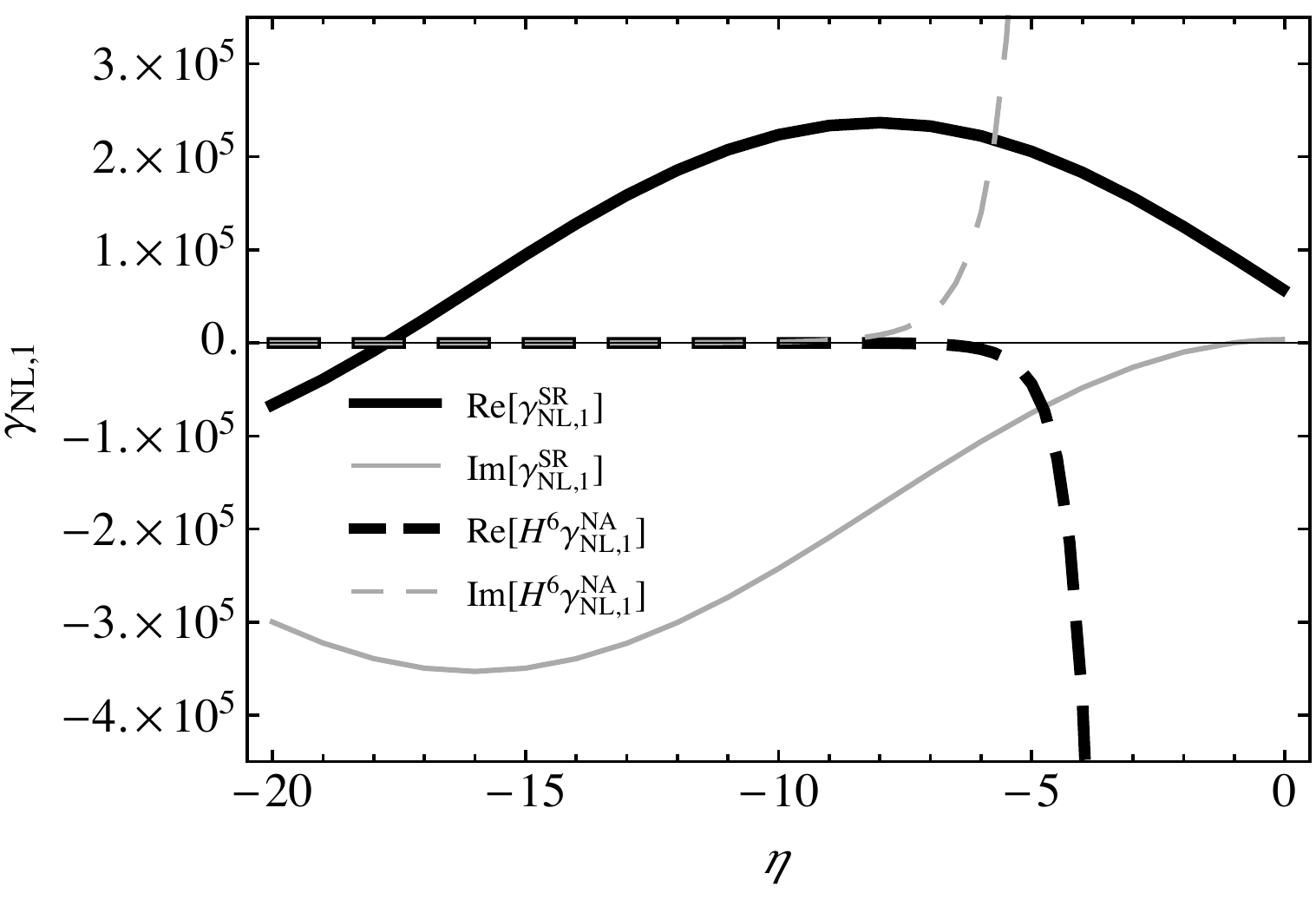}}
\caption{Example contribution to $\sum_N\hat{L}_{N1}(\eta) \hat{L}_{N2}^{\dagger}(\eta)$ from a squeezed triangle configuration with momenta (in units of $(H/c_s)$) of $k_{s1}=0.1$, $k_{s2}=0.101$ and $k_{b}=0.01$ and bath modes in (a) the quantum ground state, i.e. $m_{\vec{k}_b} = 0$, and (b) an arbitrary superposition of Fock states, i.e. summing over all possible values of $m_{\vec{k}_b}$. For both slow-roll (SR) and non-attractor (NA) dynamics we extract the dimensionless parameter $\gamma_{{\rm NL},1}$ from $q_1$ in Eq.\ (\ref{eq:DissSq}).}
\label{fig:GammasNL}
\end{figure*}

We show a more quantitative comparison of the time-dependence of these non-Hamiltonian terms between slow-roll and non-attractor models of inflation in Figs.\ \ref{fig:Gammaslin} and \ref{fig:GammasNL} for particular choices of the momentum configuration. Fig.\ \ref{fig:Gammaslin} shows the time-dependence of $\gamma_{{\rm lin},1}$ for both slow-roll and non-attractor inflation when the bath modes are in (a) the quantum ground state, i.e. $m_{\vec{k}_{b1}} = m_{\vec{k}_{b2}} = 0$, and (b) an arbitrary superposition of Fock states, i.e. summing over all occupation numbers. Fig.\ \ref{fig:GammasNL} similarly shows the time-dependence of $\gamma_{{\rm NL},1}$ when the bath modes are in the quantum ground state, i.e. $m_{\vec{k}_b} = 0$, or summed over. 

In all of the dissipation terms, there is an $\eta_1$ integral, which can be performed analytically for slow-roll squeezing parameters and numerically for the non-attractor solution. For the slow-roll case the result is that the dissipative $f_i$ and $q_i$ terms for both the folded and squeezed configurations scale like ratios of physical (not comoving) quantities times one factor that goes like the comoving momentum. The non-attractor result, however, has an additional dependence on time in the interaction strength, which makes the implications of a straightforward comparison of the numerical values between the two scenarios unclear. Rather than choose an arbitrary numerical value, we plot the quantity $H^6\gamma$. The relative qualitative time-dependence of the two cases is not affected by this choice: As the figure shows, both the linear and non-linear dissipation terms decay with time in the slow-roll case, but increase in the non-attractor case, when bath modes are summed over, as $\eta \rightarrow 0^-$. Since the non-attractor phase cannot last for more than a few e-folds, this increase at late times does not pose a problem. Further, the real parts of both the linear and nonlinear dissipation terms generically change sign as a function of time indicating that the evolution is non-Markovian \cite{Breuer:2002pc}. 
\par
%

\section{Discussion}
\label{sec:disc}

In this paper, we have presented a fully quantum framework to study the open system dynamics of inflation, with the short-long mode coupling providing the effective system-bath interaction. Our goal was to go beyond the question of standard observables and understand the full dynamics of quantum systems that have mode-coupling sample variance in their classical statistics. The results we presented take a gravitational system that has a horizon, is not static, and includes well-understood classes of interactions, and provides a bridge to less-studied quantum aspects of fields in non-trivial gravitational backgrounds. For any cubic interaction, the non-Hamiltonian terms in the evolution of modes with wavelength below some infra-red scale will be of the form shown in Eq.\ (\ref{eq:DissFolded}) for folded (two bath modes, one system mode) configurations, and Eq.\ (\ref{eq:DissSq}) for squeezed (one bath mode, two system modes) configurations. The linear and non-linear dissipation coefficients will depend on the weighted sum of triangle configurations of each type. 

The time-dependence for the two cases we considered, slow-roll and non-attractor inflation, is shown in Figs.\ \ref{fig:Gammaslin} and \ref{fig:GammasNL}, where we find a late-time growth particular to the non-attractor scenario. However, several aspects of our results are quite general: (i) for a system coupled to a long wavelength bath, folded configurations of the three-point function in momentum space lead to linear dissipation terms while squeezed configurations lead to nonlinear dissipation; (ii) since there are far fewer folded triangles with two modes in the NIR (restricting system modes in $k_{\rm min} < k_s < 2k_{\rm min}$) compared to squeezed triangles with one mode in the NIR, nonlinear dissipation is likely more significant for observable modes than linear dissipation is; (iii) ``dissipation'' does not necessarily imply the loss of coherence; indeed we find the evolution of system modes to be non-Markovian in general, irrespective of whether the bath modes are in the quantum ground state or allowed to occupy any state. Under such an evolution the system-bath interaction can lead to an exchange and even bath-mediated amplification of quantum coherences in the system. 

For the interaction we studied, the quantum environment dynamics lead to a non-Markovian system dynamics in both single-clock and non-single clock models. This implies that quantum memory of modes outside the horizon may lead to additional time-dependence of observable correlators, beyond what is uncovered in the usual semi-classical treatment. While this is not likely to be observable, we suspect the non-Markovian behavior may be especially conceptually important in understanding the quantum dynamics for non-single-clock models.

The framework presented here is appropriate for any cosmological scenario of the primordial universe where curvature modes evolve outside the horizon (or, where there is non-Gaussianity that couples modes of different wavelengths). It should facilitate a quantum open systems analysis, and decoherence studies, in the large number of non-Gaussian scenarios for which $\zeta$-correlations have already been computed, but is particularly relevant for any model with long-short mode coupling. This includes all inflation beyond single-clock, as well as contracting universe scenarios. Eventually, it may be possible to move beyond the lessons of particular models: the evolution equation we have presented here is a first step towards the appropriate effective theory \cite{Feynman:1963fq,Reiter:2012,Agon:2014uxa,Braaten:2016sja,Agon:2017oia} for observables in a large class of cosmological scenarios consistent with the current understanding of our universe. Finally, although it is unlikely that any inflationary model consistent with the classical data we have already collected will support the presence of significant late-time quantum information, this work will also facilitate exploration of whether or not such a universe is even theoretically possible.


\acknowledgments We thank P. Adshead, C. Burgess, R. Holman and A. Lawrence for useful discussions. We thank K. Sargent for suggesting improvements to (and spotting typos in) an earlier version of this paper. S. Shandera is supported by the National Science Foundation under award PHY-1719991.


\onecolumngrid

\appendix
\section{Constructing the evolution equation}
\label{app:evol}

\renewcommand{\theequation}{A\arabic{equation}}
\setcounter{equation}{0}

In this appendix we provide a few details in the derivation of Eqs.\ (\ref{eq:Inf_results})-(\ref{eq:LNs}), which give the general form of the evolution equation for the system of ``observable" modes coupled to a bath of longer wavelength, ``near infrared" modes. The full time evolution of the system is given by
\be
\hat{\sigma}(\eta)=\hat{U}(\eta,\eta_0)\hat{\sigma}(\eta_0)\hat{U}^{\dagger}(\eta,\eta_0),
\label{Eq:Densitymat}
\ee
where the time evolution operator depends on the quadratic Hamiltonian for each mode, plus any interaction term. At least for small coupling and short times, we can approximately factor out the quadratic evolution and use
\begin{subequations}
\ba
\label{eq:U(t)app}
\hat{U}(\eta,\eta_0)&=&Te^{-i\int_{\eta_0}^{\eta}\hat{H}_0(\eta_1)d\eta_1}Te^{-i\int_{\eta_0}^{\eta} \lambda(\eta_1) \hat{H}_{I,i}(\eta_1)d\eta_1} , \quad \quad \\
\hat{U}^{\dagger}(\eta,\eta_0)&=&\bar{T}e^{i \int_{\eta_0}^{\eta} \lambda(\eta_1) \hat{H}_{I,i}(\eta_1)d\eta_1} \bar{T}e^{i\int_{\eta_0}^{\eta}\hat{H}_0(\eta_1)d\eta_1} ,
\ea
\end{subequations}
where ($\bar{T}$) $T$ will (anti-) time-order the factors in the exponential. The operators in the interaction term are in the interaction picture, defined, for example, by
\be
\hat{c}_{\vec{k},i}(\eta)
=\hat{U}_{0}^{\dagger}(\eta,\eta_0)\hat{c}_{\vec{k}}(\eta_0)\hat{U}_{0}(\eta,\eta_0) \, ,
\label{Eq:intop}
\ee
where $\hat{U}_{0}(\eta,\eta_0) = T e^{-i\int_{\eta_0}^{\eta}\hat{H}_0(\eta_1)d\eta_1}$ is the propagator corresponding to the quadratic Hamiltonian. It is useful to divide the integral over momentum modes in the Fourier-space quadratic Hamiltonian at the point $k=k_{\rm min}$ (separating the observable system modes from the near infrared bath modes) and write
\be
\hat{H}_0(\eta)=\hat{H}^{\rm Obs}_0(\eta)+\hat{H}^{\rm NIR}_0(\eta) \, .
\ee
We can write the states in terms of the number of excitations for each wavenumber, using the basis of Fock states defined at $\eta_0$ for each $\vec{k}$ mode and grouped into $(\vec{k},-\vec{k})$ pairs. Then the states of all modes in the near infrared band, for example, can be written as $|N\rangle=\prod_{k\in {\rm NIR}}|m_{\vec{k}},n_{-\vec{k}}\rangle$. We assume that all modes start out in the vacuum defined at the time $\eta_0$ and denote the initial state of the set of near infrared modes as $|\psi_{\rm NIR}(\eta_0)\rangle=|\overline{NIR}\rangle$. Furthermore, since the quadratic Hamiltonian is itself time-dependent due to the presence of the two-mode squeezing term, the action of the corresponding propagator on the NIR vacuum leads to 
\bea
	\hat{U}_0(\eta,\eta_0)|0_{\vec{k}},0_{-\vec{k}}\rangle & = & \hat{S}_k(\eta)\hat{R}_k(\eta)|0_{\vec{k}},0_{-\vec{k}}\rangle \nonumber \\
	\qquad \quad & = & \frac{1}{{\rm cosh}\,r_k}\sum_{n=0}^{\infty}e^{-2in\phi_k}{\rm tanh}^nr_k|n_{\vec{k}},n_{-\vec{k}}\rangle\nonumber\\
	\qquad \quad & \equiv & |SQ(k,\eta)\rangle=\sum_n c^{{\rm sq}}_n(k,\eta)|n_{\vec{k}},n_{-\vec{k}}\rangle\,,
\eea
where $\hat{S}_k(\eta)$ and $\hat{R}_k(\eta)$ are the two-mode squeezing and rotation operators respectively, built from the time-dependent functions $r_k$ , $\phi_k$, and $\theta_{k}$. There are some different conventions for the phase $\phi$ in the literature, but notice that $\phi\rightarrow-\phi-\pi/2$ corresponds to the same squeezing angle in the quadrature plane (while changing the form of the equation above to include $(-1)^ne^{2in\phi_k}$). The squeezed state of the full bath at any given time $\eta$ can then be defined as $|SQ(\eta)\rangle=\prod_{k\in {\rm NIR}}|SQ(k,\eta)\rangle$.
\par
To find the evolution equation of observable modes we trace over the bath, comprising near infrared modes, in Eq.\ (\ref{Eq:Densitymat}) and resolve the time evolution of the reduced density matrix $\hat{\rho} (\eta) =  {\rm Tr}_{\rm NIR} \hat{\sigma}(\eta)$ at different orders of the system-bath interaction strength $\lambda (\eta)$ [introduced in Eq.\ (\ref{eq:cubicham})],
\be
\hat{\rho} (\eta) =\hat{\rho}^{(0)} (\eta) +\lambda (\eta) \hat{\rho}^{(1)}(\eta) + \lambda^{2}(\eta)\hat{\rho}^{(2)}(\eta) + \ldots \, .
\ee
Collecting terms at the lowest three orders we obtain:
\begin{itemize}
\item \underline{At lowest order} 
\begin{align}
\label{eq:order0}
\partial_{\eta}\hat{\rho}^{(0)}(\eta)
=&-i[H_0^{\rm Obs},\hat{\rho}^{(0)}(\eta)]\sum_N\left|\langle N|Te^{-i\int_{\eta_0}^{\eta}H^{\rm NIR}_0(\eta_1)d\eta_1}|\overline{NIR}\rangle\right|^2\nonumber\\
&+\hat{\rho}^{(0)}(\eta)\sum_N\left[\langle N| H_0^{\rm NIR}Te^{-i\int_{\eta_0}^{\eta}H^{\rm NIR}_0(\eta_1)d\eta_1}|\overline{NIR}\rangle\langle\overline{NIR}|\bar{T}e^{i\int_{\eta_0}^{\eta}H^{\rm NIR}_0(\eta_1)d\eta_1}|N\rangle\right.\nonumber\\
&\left.\hspace{1.8cm}-\langle N| Te^{-i\int_{\eta_0}^{\eta}H^{\rm NIR}_0(\eta_1)d\eta_1}|\overline{NIR}\rangle\langle\overline{NIR}|\bar{T}e^{i\int_{\eta_0}^{\eta}H^{\rm NIR}_0(\eta_1)d\eta_1}H_0^{\rm NIR}|N\rangle\right] .
\end{align}
The sum in the first line is equal to one, since the evolved vacuum state is normalized, and the second term, proportional to $\hat{\rho}^{(0)}$, is zero because the matrix elements are Hermitian. Then, as expected, Eq.\ (\ref{eq:order0}) simply reduces to
\be
\partial_{\eta}\hat{\rho}^{(0)}(\eta)=-i[\hat{H}^{\rm Obs}_0(\eta),\hat{\rho}^{(0)}(\eta)] \, ,
\label{Eq:rhozero}
\ee
where $\hat{\rho}^{(0)}(\eta)=Te^{-i\int_{\eta_0}^{\eta} \hat{H}^{\rm Obs}_{0}(\eta_1)d\eta_1}\hat{\rho}^{(0)}(\eta_0) \bar{T}e^{i\int_{\eta_0}^{\eta} \hat{H}^{\rm Obs}_{0}(\eta_1)d\eta_1}$.

\item \underline{At first order} 
\bea
	\partial_{\eta}\hat{\rho}^{(1)}(\eta) & = & -i\sum_N\langle N |[\hat{H}_0(\eta),\hat{\sigma}^{(1)}(t)]|N\rangle -i\left[\langle SQ(\eta)|\lambda(\eta)\hat{H}_I(\eta_0)|SQ(\eta)\rangle,\hat{\rho}^{(0)}(\eta)\right]\nonumber\\
	& = & -i\sum_N\langle N |[\hat{H}^{\rm NIR}_0(\eta),\hat{\sigma}^{(1)}(\eta)]|N\rangle -i[\hat{H}^{\rm obs}_{0}(\eta),\hat{\rho}^{(1)}(\eta)]-i\left[\langle SQ(\eta)|\lambda(\eta)H_I(\eta_0)|SQ(\eta)\rangle,\hat{\rho}^{(0)}(\eta)\right] .
\label{Eq:rholin}
\eea
Since the states $|N\rangle$ are eigenstates of only the non-squeezed part of the quadratic Hamiltonian, it may not be immediately clear that the first term in the last line vanishes. However, denoting $\hat{H}_0|N\rangle=(E_N+\hat{H}^{\rm sq}_0)|N\rangle=E_N|N\rangle +|N_{\rm sq}\rangle$ where $E_N$ is an energy, we can rewrite this term as
\bea
	\sum_N\langle N |[\hat{H}^{\rm NIR}_0(\eta),\hat{\sigma}^{(1)}(\eta)]|N\rangle & = & \sum_N\langle N |[\hat{H}_0^{\rm NIR, sq}(\eta),\hat{\sigma}^{(1)}(\eta)]|N\rangle\nonumber\\
	& = & \sum_N\left[\langle N_{sq}|\hat{\sigma}^{(1)}(\eta)|N\rangle-\langle N|\hat{\sigma}^{(1)}(\eta)|N_{sq}\rangle\right] , \qquad
\eea
which vanishes since the matrix element $\langle N_{\rm sq}|\hat{\sigma}^{(1)}(\eta)|N\rangle$ is Hermitian. (This is most easily seen by inserting a complete set of states for all NIR modes, $\sum_{N'} | N' \rangle \langle N' | = 1$, in $\langle N| \hat{H}_0^{\rm NIR}(\eta) \hat{\sigma}^{(1)}(\eta) |N\rangle$, and then using the fact that the Hamiltonian and density matrix are both Hermitian at all times.) Then, the remaining terms define an effective Hamiltonian, $\hat{H}_{\rm eff}^{(1)}=\lambda(\eta)\langle SQ(\eta)|\hat{H}_I(\eta_0)|SQ(\eta)\rangle$.

\item \underline{At second order}

We introduce the Lindblad operators,
\begin{subequations}
\bea
	\hat{L}_{N1}(\eta) & = & \langle N|\lambda(\eta)\hat{H}_I(\eta_0)|SQ(\eta)\rangle \, , \\
	\hat{L}_{N2}(\eta) & = & \int_{\eta_0}^{\eta} d\eta_1\lambda(\eta_1)\langle N|\hat{H}_{I,i}(\eta_1-\eta)|SQ(\eta)\rangle \, ,
\eea
\label{eq:LNsapp}
\end{subequations}
which allow us to write
\bea
	& & \partial_{\eta}\hat{\rho}^{(2)}(\eta) \, = \, \sum_N\left\{\hat{L}_{N1}\hat{\rho}^{(0)}(\eta)\hat{L}^{\dagger}_{N2}+\hat{L}_{N2}\hat{\rho}^{(0)}(\eta)\hat{L}^{\dagger}_{N1}-\hat{L}^{\dagger}_{N1}\hat{L}_{N2}\hat{\rho}^{(0)}(\eta)-\hat{\rho}^{(0)}(\eta)\hat{L}^{\dagger}_{N2}\hat{L}_{N1}\right\}\nonumber\\
	& & \qquad \quad \ + \, i\sum_N\bigg\{\langle N|\hat{H}_0(\eta)Te^{-i\int_{\eta_0}^{\eta}\hat{H}_0(\eta_1)d\eta_1}\int_{\eta_0}^{\eta}d\eta_1 \int_{\eta_0}^{\eta_1} d\eta_2\,\hat{H}_{I,i}(\eta_1)\hat{H}_{I,i}(\eta_2)\hat{\sigma}^{(0)}(\eta_0)\bar{T}e^{i\int_{\eta_0}^{\eta}\hat{H}_0(\eta_1)d\eta_1}|N\rangle\nonumber\\
	& & \qquad \hspace{1cm}+ \langle N|\hat{H}_0(\eta)Te^{-i\int_{\eta_0}^{\eta}\hat{H}_0(\eta_1)d\eta_1}\hat{\sigma}^{(0)}(\eta_0)\int_{\eta_0}^{\eta}d\eta_1 \int_{\eta_0}^{\eta_1} d\eta_2\,\hat{H}_{I,i}(\eta_2)\hat{H}_{I,i}(\eta_1)\bar{T}e^{i\int_{\eta_0}^{\eta}\hat{H}_0(\eta_1)d\eta_1}|N\rangle\nonumber\\
	& & \qquad \hspace{1cm}\left.- \, \langle N|\hat{H}_0(\eta)Te^{-i\int_{\eta_0}^{\eta}\hat{H}_0(\eta_1)d\eta_1}\int_{\eta_0}^{\eta}d\eta_1\hat{H}_{I,i}(\eta_1)\hat{\sigma}^{(0)}(\eta_0)\int_{\eta_0}^{\eta}d\eta_2\,\hat{H}_{I,i}(\eta_2)\bar{T}e^{i\int_{\eta_0}^{\eta}\hat{H}_0(\eta_1)d\eta_1}|N\rangle\right.\nonumber\\
	& & \qquad \quad \ + \, {\rm h.c.}\bigg\}\,.
\eea
On splitting the last term in $\{\}$ brackets into two pieces, $\eta_2<\eta_1$ and $\eta_2>\eta_1$ (and exchanging the dummy labels in the second case), it is clear that this entire term just depends on the second order density matrix,
\begin{align}
\partial_{\eta}\hat{\rho}^{(2)}(\eta)&=\sum_N\left\{\hat{L}_{N1}\hat{\rho}^{(0)}(\eta)\hat{L}^{\dagger}_{N2}+\hat{L}_{N2}\hat{\rho}^{(0)}(\eta)\hat{L}^{\dagger}_{N1}-\hat{L}^{\dagger}_{N1}\hat{L}_{N2}\hat{\rho}^{(0)}(\eta)-\hat{\rho}^{(0)}(\eta)\hat{L}^{\dagger}_{N2}\hat{L}_{N1}\right\} \nonumber\\
&\qquad -i\sum_N\langle N|[\hat{H}_0(\eta),\hat{\sigma}^{(2)}(\eta)]|N\rangle \nonumber \\
&=\sum_N\left\{\hat{L}_{N1}\hat{\rho}^{(0)}(\eta)\hat{L}^{\dagger}_{N2}+\hat{L}_{N2}\hat{\rho}^{(0)}(\eta)\hat{L}^{\dagger}_{N1}-\hat{L}^{\dagger}_{N1}\hat{L}_{N2}\hat{\rho}^{(0)}(\eta)-\hat{\rho}^{(0)}(\eta)\hat{L}^{\dagger}_{N2}\hat{L}_{N1}\right\}\nonumber\\
&\qquad -i\sum_N\langle N|[\hat{H}^{\rm Obs}_0(\eta),\hat{\rho}^{(2)}(\eta)]|N\rangle - i\sum_N\langle N|[\hat{H}^{\rm NIR}_0(\eta),\hat{\sigma}^{(2)}(\eta)]|N\rangle \, ,
\label{Eq:rhoquad}
\end{align}
where again the very last term is zero since the matrix element $\langle N_{\rm sq}|\hat{\sigma}^{(2)}(\eta)|N\rangle$ must be Hermitian. The term containing the product of Lindblad operators can further be resolved into imaginary and real contributions,
\bea
-\sum_N\hat{L}_{N1}^{\dagger}\hat{L}_{N2}\hat{\rho}^{(0)}(\eta)
&=&\sum_N\left[\frac{1}{2}\left(\hat{L}_{N2}^{\dagger}\hat{L}_{N1}-\hat{L}_{N1}^{\dagger}\hat{L}_{N2}\right) -\frac{1}{2}\left(\hat{L}_{N1}^{\dagger}\hat{L}_{N2}+\hat{L}_{N2}^{\dagger}\hat{L}_{N1}\right) \right]\hat{\rho}^{(0)}(\eta)\nonumber\\
&\equiv& [- i \hat{H}_{\rm eff}^{(2)} + \hat{A}(\eta)]\hat{\rho}^{(0)}(\eta) \, ,
\eea
by identifying the following Hermitian operators,
\begin{subequations}
\begin{align}
\hat{H}_{\rm eff}^{(2)}&=-\frac{i}{2}\sum_N(\hat{L}_{N1}^{\dagger}\hat{L}_{N2}-\hat{L}_{N2}^{\dagger}\hat{L}_{N1}) \, ,\\
\hat{A}(\eta)&=-\frac{1}{2}\sum_N \left(\hat{L}_{N1}^{\dagger}\hat{L}_{N2}+\hat{L}_{N2}^{\dagger}\hat{L}_{N1}\right).
\end{align}
\end{subequations}
\end{itemize}

Combining Eqs.\ (\ref{Eq:rhozero}), (\ref{Eq:rholin}), and (\ref{Eq:rhoquad}), we find the evolution equation reported in Eq.\ (\ref{eq:Inf_results}),
\begin{subequations}
\begin{align}
\partial_{\eta}\hat{\rho}^{(0)}(\eta) =&-i\left[\hat{H}^{\rm Obs}_{0},\hat{\rho}^{(0)}(\eta)\right] , \\
\partial_{\eta}\hat{\rho}^{(1)}(\eta) =&-i\left[\hat{H}^{\rm Obs}_{0},\hat{\rho}^{(1)}(\eta)\right]-i\left[\hat{H}_{\rm eff}^{(1)},\hat{\rho}^{(0)}(\eta)\right] , \\
\partial_{\eta}\hat{\rho}^{(2)}(\eta) =&-i\left[\hat{H}^{\rm Obs}_{0},\hat{\rho}^{(2)}(\eta)\right]-i\left[\hat{H}_{\rm eff}^{(2)},\hat{\rho}^{(0)}(\eta)\right] \nonumber \\
& \quad + \, \{\hat{A}(\eta),\hat{\rho}^{(0)}(\eta)\}+\sum_N\left[ \hat{L}_{N1}\hat{\rho}^{(0)}(\eta)\hat{L}_{N2}^{\dagger} + \hat{L}_{N2}\hat{\rho}^{(0)}(\eta)\hat{L}_{N1}^{\dagger}\right] .
\end{align}
\label{eq:Inf_resultsapp}
\end{subequations}


\section{Lindblad terms from a $\zeta\dot{\zeta}^2$ interaction}
\label{app:lind}

\renewcommand{\theequation}{B\arabic{equation}}
\setcounter{equation}{0}

In this appendix we show how we evaluate terms in the Lindbladian, such as $\sum_N\hat{L}_{N1}(\eta)\hat{\rho}^{(0)}(\eta)\hat{L}^{\dagger}_{N2}(\eta)$, given a system-bath interaction. The specific interaction we consider is the cubic action for the curvature perturbation, $S_3=M_p^2\int d^3x \,d\eta\, a^4(3\epsilon/c_s^2)(c_s^2-1)\zeta\dot{\zeta}^2$. In Fourier space and in terms of creation and annihilation operators of the canonical field $\chi$, this leads to the interaction Hamiltonian written in Eq.\ (\ref{eq:cubicham}),
\ba
\lambda(\eta)\hat{H}_I&=&\frac{3(c_s^2-1)}{8M_pc_s^2a\sqrt{\epsilon}}\int_{\triangle} \[ \sqrt{\frac{k_2k_3}{k_1}}\left(\hat{c}^{\dagger}_{-\vec{k}_1}\hat{c}^{\dagger}_{-\vec{k}_2}\hat{c}^{\dagger}_{-\vec{k}_3}+\hat{c}_{\vec{k}_1}\hat{c}^{\dagger}_{-\vec{k}_2}\hat{c}^{\dagger}_{-\vec{k}_3}+ \ldots \right) +{\rm perm.} \] ,
\label{Eq:Hint}
\ea
where we have used the shorthand $\int_{\triangle}=\int \frac{d^3k_1}{(2\pi)^3} \frac{d^3k_2}{(2\pi)^3} \frac{d^3k_3}{(2\pi)^3}(2\pi)^3\delta^3(\vec{k}_1+\vec{k}_2+\vec{k}_3).$ The terms inside the parenthesis include all possible momentum conserving combinations of operators, with some terms appearing with a minus sign since the interaction term couples the field $\chi$ and its conjugate momentum. The pre-factors of the integral define a dimensionless, but time-dependent coupling coefficient ${\lambda(\eta)=3(c_s^2-1)/(8c_s^2a(\eta)\sqrt{\epsilon(\eta)})}$. Using the fact that $\epsilon(\eta)$ is approximately constant for slow-roll and $\sim a^{-6}(\eta)$ for non-attractor models, we obtain the following expressions for the coupling,
\begin{subequations}
\begin{align}
\lambda^{\rm SR}(\eta)=&-\frac{3(c_s^2-1)}{8c_s^2\sqrt{\epsilon}}\eta H \, , \\
\lambda^{\rm NA}(\eta)=&\frac{3}{8}\frac{(c_s^2-1)}{c_s^2}\left(\frac{1}{H\eta}\right)^2.
\end{align}
\label{eq:lambdas}
\end{subequations}
\par
For the cubic interaction, we consider cases where one, two, or three of the momenta are bath modes, i.e., they belong to the NIR band. Notice that since the terms that depend on $\hat{L}_{Ni}$ always come with $\sum_N$, they will give non-zero contributions only when the same number of modes are in the NIR in both $\hat{L}_{Ni}$ and $\hat{L}^{\dagger}_{Nj}$. That, in turn, means that there will always be an even number of $\hat{c}_{\vec{k}}$, $\hat{c}^{\dagger}_{\vec{k}}$ operators for modes in the observable band. As the first, trivial case, suppose all three momenta are in the NIR. Then the $\hat{L}_{Ni}$ are just numbers and so $\hat{H}_{\rm eff}^{(2)}=0$ and the terms in the last line of Eq.\ (\ref{eq:Inf_resultsapp}) all sum to zero. 
\par
It is helpful to write the interaction Hamiltonian for the two other cases:
\begin{enumerate}[(i)]
\item  ``folded" triangles with two NIR modes and one observable mode, and 
\item ``squeezed" triangles with one NIR mode and two observable modes. 
\end{enumerate}
Writing
\begin{align}
\lambda(\eta)\hat{H}_I=\frac{\lambda(\eta)}{M_p}\int_{\triangle}\hat{F}(k_1,k_2,k_3) \, ,
\end{align}
we can write the function $\hat{F}$ for the folded triangle case as
\begin{align}
\hat{F}^{\rm fold}=\,&\hat{c}_{\vec{k}_s}\left\{\sqrt{\frac{k_{b1}k_{b2}}{k_s}}\left[\hat{c}_{\vec{k}_{b1}}\hat{c}_{\vec{k}_{b2}}+\hat{c}^{\dagger}_{-\vec{k}_{b1}}\hat{c}^{\dagger}_{-\vec{k}_{b2}}-\hat{c}_{\vec{k}_{b1}}\hat{c}^{\dagger}_{-\vec{k}_{b2}}-\hat{c}^{\dagger}_{-\vec{k}_{b1}}\hat{c}_{\vec{k}_{b2}}\right]+\sqrt{\frac{k_sk_{b2}}{k_{b1}}}\left[+--+\right]+\sqrt{\frac{k_sk_{b1}}{k_{b2}}}\left[+-+-\right]\right\}\nonumber\\
&+\hat{c}^{\dagger}_{-\vec{k}_s}\left\{\sqrt{\frac{k_{b1}k_{b2}}{k_s}}\left[++--\right]+\sqrt{\frac{k_sk_{b2}}{k_{b1}}}\left[-++-\right]+\sqrt{\frac{k_sk_{b1}}{k_{b2}}}\left[-+-+\right]\right\},
\end{align}
where $k_1\equiv k_s$ and $k_{b1}$, $k_{b2}$ denote the momenta associated with the system (observable modes) and NIR modes respectively. Here $+$, $-$ are a shorthand for the appropriately signed sum of the same combinations of operators as in the first set of square brackets.
\par
Similarly, for squeezed triangles, 
\begin{align}
\hat{F}^{\rm sq}=\,& \hat{c}_{\vec{k}_{s1}}\hat{c}_{\vec{k}_{s2}}\left[\sqrt{\frac{k_{s1}k_{s2}}{k_b}}\left(\hat{c}_{\vec{k}_b}+\hat{c}^{\dagger}_{-\vec{k}_b}\right)+\sqrt{\frac{k_bk_{s2}}{k_{s1}}}\,(+-)+\sqrt{\frac{k_{s1}k_b}{k_{s2}}}\,(+-)\right]\nonumber\\
& +\hat{c}^{\dagger}_{-\vec{k}_{s1}}\hat{c}^{\dagger}_{-\vec{k}_{s2}}\left[\sqrt{\frac{k_{s1}k_{s2}}{k_b}}\left(++\right)+\sqrt{\frac{k_bk_{s2}}{k_{s1}}}\,(-+)+\sqrt{\frac{k_{s1}k_b}{k_{s2}}}\,(-+)\right]\nonumber\\
&+ \hat{c}_{\vec{k}_{s1}}\hat{c}^{\dagger}_{-\vec{k}_{s2}}\left[\sqrt{\frac{k_{s1}k_{s2}}{k_b}}\left(--\right)+\sqrt{\frac{k_bk_{s2}}{k_{s1}}}\,(-+)+\sqrt{\frac{k_{s1}k_b}{k_{s2}}}\,(+-)\right]\nonumber\\
&+ \hat{c}^{\dagger}_{-\vec{k}_{s1}}\hat{c}_{-\vec{k}_{s2}}\left[\sqrt{\frac{k_{s1}k_{s2}}{k_b}}\left(--\right)+\sqrt{\frac{k_bk_{s2}}{k_{s1}}}\,(+-)+\sqrt{\frac{k_{s1}k_b}{k_{s2}}}\,(-+)\right],
\end{align}
where momenta $k_{s1}$, $k_{s2}$ identify the observable modes, and the momentum $k_b$ is associated with the NIR mode. 
%
\subsection{$\hat{L}_{N1}$, $\hat{L}_{N2}$ for folded triangles}
%
For folded triangle configurations, involving one system mode, $k_{s}$, and two bath modes, $k_{b1}$, $k_{b2}$, we find that
\begin{align}
\label{eq:LN1fold}
\hat{L}^{\rm fold}_{N1}(\eta)=&\frac{\lambda(\eta)}{M_p}\int_{\triangle}\frac{1}{(k_{b1}k_{b2})^{3/2}}\:\prod_{k_i\in {\rm NIR}, k_i\neq k_{b1}, k_{b2}}\langle m_{\vec{k}_i},n_{-\vec{k}_i}|SQ(k,\eta)\rangle\,\delta_{m_{\vec{k}_{b1}}+1,n_{\vec{k}_{b1}}}\delta_{m_{\vec{k}_{b2}}+1,n_{\vec{k}_{b2}}}\sqrt{(m_{\vec{k}_{b1}}+1)(m_{\vec{k}_{b2}}+1)}\nonumber\\
&\times\left\{\hat{c}_{\vec{k}_o}(\eta_0)\left[\sqrt{\frac{k_{b1}k_{b2}}{k_s}}\left[c^{\rm sq}_{n_{\vec{k}_{b1}}}(k_{b1},\eta)c^{\rm sq}_{n_{\vec{k}_{b2}}}(k_{b2},\eta)+c^{\rm sq}_{m_{\vec{k}_{b1}}}c^{\rm sq}_{m_{\vec{k}_{b2}}}-c^{\rm sq}_{n_{\vec{k}_{b1}}}c^{\rm sq}_{m_{\vec{k}_{b2}}}-c^{\rm sq}_{m_{\vec{k}_{b1}}}c^{\rm sq}_{n_{\vec{k}_{b2}}}\right]\right.\right.\nonumber\\
&\hspace{1.5cm}\left.+\sqrt{\frac{k_sk_{b2}}{k_{b1}}}\,[+--+]+\sqrt{\frac{k_sk_{b1}}{k_{b2}}}\,[+-+-]\right]\nonumber\\
&\hspace{0.8cm}+\left.\hat{c}^{\dagger}_{-\vec{k}_o}(\eta_0)\left[\sqrt{\frac{k_{b1}k_{b2}}{k_s}}[++--]+\sqrt{\frac{k_sk_{b2}}{k_{b1}}}\,[-++-]+\sqrt{\frac{k_sk_{b1}}{k_{b2}}}\,[-+-+]\right]\right\},
\end{align}
where we have used $\hat{c}_{\vec{k}}(\eta_0)|n_{\vec{k}}\rangle = \frac{\sqrt{n}}{k^{3/2}} |(n-1)_{\vec{k}}\rangle$; the factor of $k^{-3/2}$ here is consistent with the commutation relation that tells us that our ladder operators have dimensions of $k^{-3/2}$. 
\par
To evaluate $\hat{L}_{N2}$, we use the interaction picture representation of operators from Eq.\ (\ref{Eq:intop}),
\begin{align}
\hat{c}_{\vec{k},i}(\eta)
=\hat{U}_{0}^{\dagger}(\eta,\eta_0)\hat{c}_{\vec{k}}(\eta_0)\hat{U}_{0}(\eta,\eta_0)
&=\prod_{k^{\prime}}e^{i\int_{\eta_0}^{\eta}\hat{H}_0(\vec{k}^{\prime},\eta_1)d\eta_1}\hat{c}_{\vec{k}}(\eta_0)\prod_{k^{\prime\prime}}e^{-i\int_{\eta_0}^{\eta}\hat{H}_0(\vec{k}^{\prime\prime},\eta_1)d\eta_1}\nonumber\\
&=e^{i\int_{\eta_0}^{\eta}\hat{H}_0(\vec{k},\eta_1)d\eta_1}\hat{c}_{\vec{k}}(\eta_0)e^{-i\int_{\eta_0}^{\eta}\hat{H}_0(\vec{k},\eta_1)d\eta_1}\nonumber\\
&=\hat{R}^{\dagger}_k(\eta)\hat{S}^{\dagger}_k(\eta)\hat{c}_{\vec{k}}(\eta_0)\hat{S}_k(\eta)\hat{R}_k(\eta)\nonumber\\
&=u_{k}(\eta)\hat{c}_{\vec{k}}(\eta_0)+v_{k}(\eta)\hat{c}_{-\vec{k}}^{\dagger}(\eta_0),
\end{align}
and we denote
\begin{subequations}
\begin{align}
\hat{c}_{\vec{k},i}(\eta_{1}-\eta) 
&= \hat{U}_{0}(\eta,\eta_0)\hat{U}_{0}^{\dagger}(\eta_{1},\eta_0) \hat{c}_{\vec{k}}(\eta_0) \hat{U}_{0}(\eta_{1},\eta_0)  \hat{U}_{0}^{\dagger}(\eta,\eta_0)\nonumber\\
&=\hat{S}_k(\eta)\hat{R}_k(\eta)\hat{R}^{\dagger}_k(\eta_1)\hat{S}^{\dagger}_k(\eta_1)\hat{c}_{\vec{k}}(\eta_0)\hat{S}_k(\eta_1)\hat{R}_k(\eta_1)\hat{R}^{\dagger}_k(\eta)\hat{S}^{\dagger}_k(\eta) \nonumber\\
&=\hat{S}_k(\eta)\hat{R}_k(\eta)
\left[u_{k}(\eta_1)\hat{c}_{\vec{k}}(\eta_0)+v_{k}(\eta_1)\hat{c}^{\dagger}_{-\vec{k}}(\eta_0)\right]\hat{R}^{\dagger}_k(\eta)\hat{S}^{\dagger}_k(\eta) \, , \\
\hat{c}^{\dagger}_{-\vec{k},i}(\eta_{1}-\eta) &=\hat{S}_k(\eta)\hat{R}_k(\eta)\left[u_{k}^*(\eta_1)\hat{c}^{\dagger}_{-\vec{k}}(\eta_0)+v_{k}^*(\eta_1)\hat{c}_{\vec{k}}(\eta_0)\right]\hat{R}^{\dagger}_k(\eta)\hat{S}^{\dagger}_k(\eta) \, ,
\end{align}
\end{subequations}
where $u$ and $v$ are the complex functions described in section\ \ref{sec:srna}.
Substituting the interaction picture operators in the expression for $\hat{L}_{N2}$ [Eq.\ (\ref{eq:LNsapp})], we obtain
\begin{align}
\label{eq:LN2fold}
\hat{L}^{\rm fold}_{N2}(\eta)=&\frac{1}{M_p}\int_{\triangle}\frac{1}{(k_{b1}k_{b2})^{3/2}}\:\prod_{k_i\in {\rm NIR}, k_i\neq k_{b1}, k_{b2}}\langle m_{\vec{k}_i},n_{-\vec{k}_i}|SQ(k,\eta)\rangle\,\delta_{m_{\vec{k}_{b1}}+1,n_{\vec{k}_{b1}}}\delta_{m_{\vec{k}_{b2}}+1,n_{\vec{k}_{b2}}}\nonumber\\
&\times\hat{S}_{k_s}(\eta)\hat{R}_{k_s}(\eta)\frac{e^{-i\theta_{k_{b1}}(\eta)}}{{\rm cosh}\,r_{k_{b1}}(\eta)}\frac{e^{-i\theta_{k_{b2}}(\eta)}}{{\rm cosh}\,r_{k_{b2}}(\eta)}\sqrt{(m_{\vec{k}_{b1}}+1)(m_{\vec{k}_{b2}}+1)}\;c^{\rm sq}_{m_{\vec{k}_{b1}}}(k_{b1},\eta)c^{\rm sq}_{m_{\vec{k}_{b2}}}(k_{b2},\eta)\nonumber\\
&\times\Bigg[\hat{c}_{\vec{k}_s}(\eta_0)\hat{R}^{\dagger}_{k_s}(\eta)\hat{S}^{\dagger}_{k_s}(\eta)\int_{\eta_0}^{\eta}d\eta_1\lambda(\eta_1)\left\{\sqrt{\frac{k_{b1}k_{b2}}{k_s}}\left[u_{k_s}(\eta_1)\,\left(v_{k_{b1}}(\eta_1)v_{k_{b2}}(\eta_1)+u^*u^*-vu^*-u^*v\right)\right.\right.\nonumber\\
&\hspace{6.6cm}\left.+ \, v^*_{k_s}(\eta_1)(++--)\right]\nonumber\\
&\hspace{6.6cm}+\sqrt{\frac{k_sk_{b2}}{k_{b1}}}\left[u_{k_s}(\eta_1)(+--+)+v^*_{k_s}(\eta_1)(-++-)\right]\nonumber\\
&\hspace{6.6cm}\left.+\sqrt{\frac{k_{b1}k_s}{k_{b2}}}\left[u_{k_s}(\eta_1)(+-+-)+v^*_{k_s}(\eta_1)(-+-+)\right]\right\}\nonumber\\
&\hspace{0.6cm}+\hat{c}^{\dagger}_{-\vec{k}_s}(\eta_0)\hat{R}^{\dagger}_{k_s}(\eta)\hat{S}^{\dagger}_{k_s}(\eta)\int_{\eta_0}^{\eta}d\eta_1\lambda(\eta_1)\left\{\sqrt{\frac{k_{b1}k_{b2}}{k_s}}\left[v_{k_s}(\eta_1)(++--)+u^*_{k_s}(\eta_1)(++--)\right]\,\right.\nonumber\\
&\hspace{6.6cm}+\sqrt{\frac{k_sk_{b2}}{k_{b1}}}\left[v_{k_s}(\eta_1)(+--+)+u^*_{k_s}(\eta_1)(-++-)\right]\nonumber\\
&\hspace{6.6cm}\left.+\sqrt{\frac{k_{b1}k_s}{k_{b2}}}\left[v_{k_s}(\eta_1)(+-+-)+u^*_{k_s}(\eta_1)(-+-+)\right]\right\}\Bigg] ,
\end{align}
where we have also used the results
\begin{align}
&  \hat{S}_k(\eta)\hat{R}_k(\eta)|0_{\vec{k}},0_{-\vec{k}}\rangle = |SQ(k,\eta)\rangle = \sum_n c^{{\rm sq}}_n(k,\eta)|n_{\vec{k}},n_{-\vec{k}}\rangle\,,\\
&  \hat{S}_k(\eta)\hat{R}_k(\eta)|0_{\vec{k}},1_{-\vec{k}}\rangle=\frac{e^{-i\theta_k(\eta)}}{{\rm cosh}\,r_k(\eta)}\sum_n\sqrt{n+1}\;c^{\rm sq}_n(k,\eta)|n_{\vec{k}},(n+1)_{-\vec{k}}\rangle\,.
\end{align}
%
%
\subsection{$\hat{L}_{N1}$, $\hat{L}_{N2}$ for squeezed triangles}
%
Following similar steps as in the folded case, we obtain the following expressions for the Lindblad operators for the squeezed configuration of two system modes, $k_{{s}1}$, $k_{{s}2}$, and one bath mode, $k_b$,
\begin{align}
\label{eq:LN1sq}
\hat{L}^{\rm sq}_{N1}(\eta)=&\frac{\lambda(\eta)}{M_p}\int_{\triangle}\frac{1}{(k_b)^{3/2}}\:\prod_{k_i\in {\rm NIR}, k_i\neq k_b}\langle m_{\vec{k}_i},n_{-\vec{k}_i}|SQ(k,\eta)\rangle\,\delta_{m_{\vec{k}_b}+1,n_{\vec{k}_b}}\sqrt{m_{\vec{k}_b}+1}\nonumber\\
&\times\left\{\hat{c}_{\vec{k}_{s1}}(\eta_0)\hat{c}_{\vec{k}_{s2}}(\eta_0)\left[\sqrt{\frac{k_{s1}k_{s2}}{k_b}}\left[c^{\rm sq}_{n_{\vec{k}_b}}(k_b,\eta)+c^{\rm sq}_{m_{\vec{k}_b}}(k_b,\eta)\right]+\sqrt{\frac{k_bk_{s2}}{k_{s1}}}\,[+-]+\sqrt{\frac{k_bk_{s1}}{k_{s2}}}\,[+-]\right]\right.\nonumber\\
&\hspace{1cm}+\hat{c}^{\dagger}_{-\vec{k}_{s1}}(\eta_0)\hat{c}^{\dagger}_{-\vec{k}_{s2}}(\eta_0)\left[\sqrt{\frac{k_{s1}k_{s2}}{k_b}}\,[++]+\sqrt{\frac{k_bk_{s2}}{k_{s1}}}\,[-+]+\sqrt{\frac{k_bk_{s1}}{k_{s2}}}\,[-+]\right]\nonumber\\
&\hspace{1cm}+\hat{c}_{\vec{k}_{s1}}(\eta_0)\hat{c}^{\dagger}_{-\vec{k}_{s2}}(\eta_0)\left[\sqrt{\frac{k_{s1}k_{s2}}{k_b}}\,[--]+\sqrt{\frac{k_bk_{s2}}{k_{s1}}}\,[-+]+\sqrt{\frac{k_bk_{s1}}{k_{s2}}}\,[+-]\right]\nonumber\\
&\hspace{1cm}\left.+\,\hat{c}^{\dagger}_{-\vec{k}_{s1}}(\eta_0)\hat{c}_{\vec{k}_{s2}}(\eta_0)\left[\sqrt{\frac{k_{s1}k_{s2}}{k_b}}\,[--]+\sqrt{\frac{k_bk_{s2}}{k_{s1}}}\,[+-]+\sqrt{\frac{k_bk_{s1}}{k_{s2}}}\,[-+]\right]\right\} ,
\end{align}
\begin{align}
\label{eq:LN2sq}
&\hat{L}^{\rm sq}_{N2}(\eta) \nonumber\\
&\ \ =\frac{1}{M_p}\int_{\triangle}\frac{1}{(k_b)^{3/2}}\:\prod_{k_i\in {\rm NIR}, k_i\neq k_b}\langle m_{\vec{k}_i},n_{-\vec{k}_i}|SQ(k,\eta)\rangle\,\delta_{m_{\vec{k}_b}+1,n_{\vec{k}_b}}\frac{e^{-i\theta_{k_b}(\eta)}}{{\rm cosh}\,r_{k_b}(\eta)}\sqrt{(m_{\vec{k}_b}+1)}\;c^{\rm sq}_{m_{\vec{k}_b}}(k_b,\eta)\nonumber\\
&\hspace{0.8cm}\times\hat{S}_{k_{{s}1}}(\eta)\hat{R}_{k_{{s}1}}(\eta)\hat{S}_{k_{{s}2}}(\eta)\hat{R}_{k_{{s}2}}(\eta)
\Bigg\{\hat{c}_{\vec{k}_{{s}1}}(\eta_0)\hat{c}_{\vec{k}_{{s}2}}(\eta_0)\nonumber\\
& \hspace{0.8cm}\times\int_{\eta_0}^{\eta} d\eta_1\lambda(\eta_1)\left\{u_{k_{s1}}(\eta_1)u_{k_{s2}}(\eta_1)\left[\sqrt{\frac{k_{{s}1}k_{{s}2}}{k_b}}\,[v_{k_b}(\eta_1)+u^*_{k_b}(\eta_1)]+ \sqrt{\frac{k_b k_{{s}2}}{k_{{s}1}}}(+-)+\sqrt{\frac{k_{{s}1}k_b}{k_{{s}2}}}(+-) \right] \right. \nonumber\\
&\hspace{4cm}+v^*_{k_{s1}}(\eta_1)v^*_{k_{s2}}(\eta_1)\left[\sqrt{\frac{k_{{s}1}k_{{s}2}}{k_b}}(++)+\sqrt{\frac{k_bk_{{s}2}}{k_{{s}1}}}(-+)+\sqrt{\frac{k_{{s}1}k_b}{k_{{s}2}}}(-+)\right]\nonumber\\
&\hspace{4cm}+u_{k_{s1}}(\eta_1)v^*_{k_{s2}}(\eta_1)\left[\sqrt{\frac{k_{{s}1}k_{{s}2}}{k_b}}(--)+\sqrt{\frac{k_b k_{{s}2}}{k_{{s}1}}}(-+)+\sqrt{\frac{k_{{s}1}k_b}{k_{{s}2}}}(+-)\right]\nonumber\\
&\left.\hspace{4cm}+ \, v^*_{k_{s1}}(\eta_1)u_{k_{s2}}(\eta_1)\left[\sqrt{\frac{k_{{s}1}k_{{s}2}}{k_b}}(--)+\sqrt{\frac{k_bk_{{s}2}}{k_{{s}1}}}(+-)+\sqrt{\frac{k_{{s}1}k_b}{k_{{s}2}}}(-+)\right]\right\}\nonumber\\
&\hspace{0.8cm}+\hat{c}_{\vec{k}_{{s}1}}(\eta_0)\hat{c}^{\dagger}_{-\vec{k}_{{s}2}}(\eta_0)\int_{\eta_0}^{\eta}d\eta_1\lambda(\eta_1)\left\{u_{k_{s1}}(\eta_1)v_{k_{s2}}(\eta_1)[(1)]+v^*u^*[(2)]+uu^*[(3)]+v^*v[(4)]\right\}\nonumber\\
&\hspace{0.8cm}+\hat{c}^{\dagger}_{-\vec{k}_{{s}1}}(\eta_0)\hat{c}_{\vec{k}_{{s}2}}(\eta_0)\int_{\eta_0}^{\eta}d\eta_1\lambda(\eta_1)\left\{vu[(1)]+u^*v^*[(2)]+vv^*[(3)]+u^*u[(4)]\right\}\nonumber\\
&\hspace{0.8cm}+\hat{c}^{\dagger}_{-\vec{k}_{{s}1}}(\eta_0)\hat{c}^{\dagger}_{-\vec{k}_{{s}2}}(\eta_0)\int_{\eta_0}^{\eta}d\eta_1\lambda(\eta_1)\left\{vv[(1)]+u^*u^*[(2)]+vu^*[(3)]+u^*v[(4)]\right\}\Bigg\}\hat{R}^{\dagger}_{k_{{s}1}}(\eta)\hat{S}^{\dagger}_{k_{{s}1}}(\eta)\hat{R}^{\dagger}_{k_{{s}2}}(\eta)\hat{S}^{\dagger}_{k_{{s}2}}(\eta)\,.
\end{align}

\end{document}